%% file: main.tex
\documentclass[sigconf]{acmart}

\AtBeginDocument{%
  \providecommand\BibTeX{{%
    \normalfont B\kern-0.5em{\scshape i\kern-0.25em b}\kern-0.8em\TeX}}}

\copyrightyear{2024}
\acmYear{2024}
\setcopyright{rightsretained}
\acmConference[IUI '24]{29th International Conference on Intelligent User Interfaces}{March 18--21, 2024}{Greenville, SC, USA}
\acmBooktitle{29th International Conference on Intelligent User Interfaces (IUI '24), March 18--21, 2024, Greenville, SC, USA}
\acmDOI{10.1145/3640543.3645208}
\acmISBN{979-8-4007-0508-3/24/03}

\usepackage{algorithm}
\usepackage{algpseudocode} 
\usepackage{xcolor}
\usepackage{wrapfig}
\usepackage{hyperref}
\usepackage{listings}
\usepackage[frozencache,cachedir=.]{minted}

\definecolor{chart_red}{HTML}{EC5900}
\definecolor{chart_blue}{HTML}{0066b2}
\definecolor{chart_green}{HTML}{00A86B}

\newcommand{\toolname}[0]{\textsc{SlopeSeeker}}

\newcommand{\pheading}[1]{\vspace{4px}\noindent\textbf{#1}}

\acmSubmissionID{8501}

\begin{document}

\title{\toolname{}: A Search Tool for Exploring a Dataset of Quantifiable Trends} 

\author{Alexander Bendeck}
\email{abendeck3@gatech.edu}
\orcid{0000-0002-9799-2194}
\affiliation{%
  \institution{Tableau Research, Georgia~Institute~of~Technology}
  \city{Atlanta}
  \state{Georgia}
  \country{USA}
}

\author{Dennis Bromley}
\email{dbromley@tableau.com}
\orcid{0009-0007-0303-8062}
\affiliation{%
  \institution{Tableau Research}
  \city{Seattle}
  \state{Washington}
  \country{USA}
}

\author{Vidya Setlur}
\email{vsetlur@tableau.com}
\orcid{0000-0003-3722-406X}
\affiliation{%
  \institution{Tableau Research}
  \city{Palo Alto}
  \state{California}
  \country{USA}
}

\renewcommand{\shortauthors}{Bendeck, Bromley, and Setlur}

\begin{abstract}
\input{sections/0-abstract}
\end{abstract}



\begin{teaserfigure}
  \includegraphics[width=\textwidth]{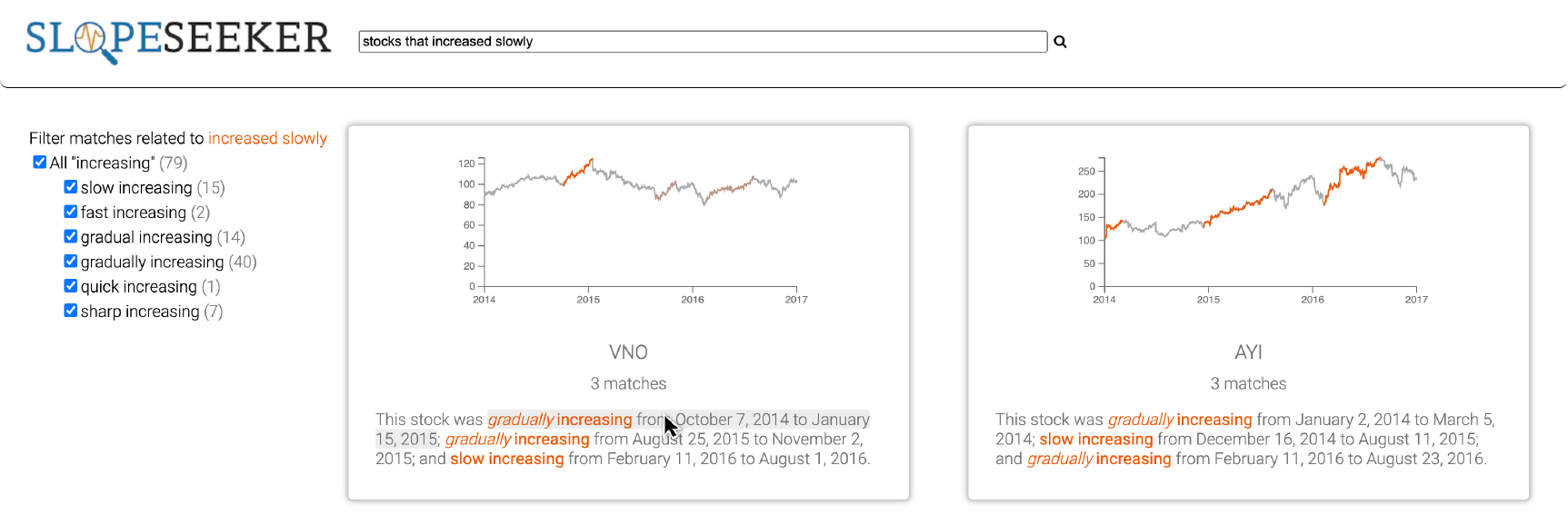}
  \caption{The \toolname{} interface.  At the top, the search bar indicates a search for ``stocks that increased slowly.'' Below that, the interface includes (from left to right) filter checkboxes showing labels related to ``increased slowly'' and two search result tiles, each showing a different stock price over time. The user is hovering over a text annotation in the left result tile. As a result, the tile's stock-description text is highlighted in \textcolor{gray}{gray}, and the corresponding data trend on the stock price chart above is emphasized in \textcolor{chart_red}{red} while other data trends on the chart appear faded.}
  \Description[An image of the SlopeSeeker interface.]{An image of the SlopeSeeker interface. Fully described in the main figure text.}
  \label{fig:teaser}
\end{teaserfigure}


\maketitle

\input{sections/1-introduction}

\input{sections/2-related-work}

\input{sections/3-experiments-data}

\input{sections/4-slopeseeker}

\input{sections/5-evaluation}

\input{sections/6-future-work}

\input{sections/7-conclusion}


\bibliographystyle{ACM-Reference-Format}
\bibliography{main}

\end{document}

%% file: sections/0-abstract.tex
Natural language and search interfaces intuitively facilitate data exploration and provide visualization responses to diverse analytical queries based on the underlying datasets. However, these interfaces often fail to interpret more complex analytical intents, such as discerning subtleties and quantifiable differences between terms like ``bump’’ and ``spike’’ in the context of COVID cases, for example. We address this gap by extending the capabilities of a data exploration search interface for interpreting semantic concepts in time series trends. We first create a comprehensive dataset of semantic concepts by mapping quantifiable univariate data trends such as \textit{slope} and \textit{angle} to crowdsourced, semantically meaningful trend labels. The dataset contains quantifiable properties that capture the slope-scalar effect of semantic modifiers like ``sharply'' and ``gradually,'' as well as multi-line trends (e.g., ``peak,'' ``valley''). We demonstrate the utility of this dataset in \toolname{}, a tool that supports natural language querying of quantifiable trends, such as ``\textit{show me stocks that tanked in 2010}.'' The tool incorporates novel scoring and ranking techniques based on semantic relevance and visual prominence to present relevant trend chart responses containing these semantic trend concepts. In addition, \toolname{} provides a faceted search interface for users to navigate a semantic hierarchy of concepts from general trends (e.g., ``increase’’) to more specific ones (e.g., ``sharp increase’’). A preliminary user evaluation of the tool demonstrates that the search interface supports greater expressivity of queries containing concepts that describe data trends. We identify potential future directions for leveraging our publicly available quantitative semantics dataset in other data domains and for novel visual analytics interfaces.

%% file: sections/1-introduction.tex
\section{Introduction}
Trends are patterns in data that indicate the general change in an attribute with time~\cite{chatfield2004timeseries}. Searching for trends over time is a prevalent task in data analysis tools to identify anomalies or deviations from the normal or expected values in a dataset~\cite{tableau,powerbi,soukup2002visual,de2003visual,kincaid2006line}. Trend analysis has significant relevance across application domains, ranging from discerning stock market trajectories and economic fluctuations to studying climate patterns, urban growth patterns, and monitoring disease epidemiology and health behavior~\cite{tunnicliffe:2016}.

During the height of the COVID-19 pandemic, for example, trends in the number of confirmed virus cases were constantly analyzed and compared between different geographic regions and time frames to understand the nature of the virus and the impact of different public health policies and mitigation measures. Enabling users to search for these types of trends using natural language (NL) would allow them great power and flexibility to express their intents. The difference between a ``slow increase'' and a ``rapid increase'' in COVID-19 cases could have huge implications for public policy; similarly, the difference between a stock price ``slumping'' and ``tanking'' is likely to invoke drastically different investment decisions. The expressive power in these scenarios comes from the precise, quantified semantics of these words used to describe the trends, and we argue that there is an opportunity for search tools to leverage these semantics to interpret expressive user analytical intents. Specifically, a set of quantifiable semantic labels can provide useful metadata to necessitate a structured approach to indexing, classification, and retrieval of trends in a search tool. Metadata that can encapsulate language that describes slopes and angles can further enhance the precision and recall of trends in search tools. 

Search tools have recently evolved beyond document search, supporting intents for data exploration and providing results that include visualizations or widgets displaying data relevant to the user's query~\cite{olio}. 
Similarly, natural language interfaces (NLIs) for visual data analysis~\cite{eviza,datatone,thoughtspot,ibmwatson,setlur2019inferencing,analyza} now enable users to engage with and query data using NL. However, both these search tools and NLIs support only basic analytical intents, with limited support for interpreting temporal trends~\cite{amar2005tasks}. 

\textbf{Contributions.} In this work, we explore the potential of allowing users to search for temporal phenomena in a dataset by leveraging precise, quantified semantics of language, focusing on searching for trends in time series data. Specifically, our contributions are as follows: 
\begin{itemize}
    \item We collect a comprehensive dataset of semantic concepts describing trends and their quantifiable properties through crowdsourced data collection experiments. Going beyond prior work in this space~\cite{bromley2023difference}, our dataset maps numeric slopes to semantic trend descriptor words and phrases; for example, we include slope labels (e.g., ``falling'') and slope labels with modifiers (``slowly falling''), along with multi-line trends that comprise a combination of ``up,'' ``down,'' and ``flat'' trend segments (e.g., ``peak,'' ``valley''). We release the quantified semantic trends dataset publicly\footnote{Link: \href{https://osf.io/yzdvt/?view\_only=d3723224f9234776a10882eee8b7568a}{https://osf.io/yzdvt/?view\_only=d3723224f9234776a10882eee8b7568a}}. Based on this dataset, we also introduce an approach for applying semantic trend descriptor labels to raw time series data. 
    \item To demonstrate the applicability of our semantic trend labels dataset, we present the \toolname{} tool, which implements a novel analytical search experience supporting diverse trend search intents for these labeled trends. \toolname{} incorporates novel scoring and ranking techniques based on both the label relevance and visual prominence of trends. The tool also surfaces a semantic hierarchy of trend descriptor terms from our dataset, with which the user can interact to filter results.
    \item Using the \toolname{} tool as a design probe, we conduct a qualitative study with $12$ participants to gain feedback on the trend querying features in the tool, the system design and implementation behavior, and how the labeled dataset aids in returning relevant trend search results. The study verifies that the trend-focused search paradigm effectively supports the distinct objectives of searching for pre-identified trend patterns. Finally, drawing from our observational data and participant responses, we identify potential directions for further development of the tool and the underlying semantic labeled dataset. 
    \item We identify and discuss promising directions for future work in this space, such as employing LLMs for data augmentation and generating trend narratives, as well as exploring time normalization and predictive analytics for more nuanced interpretations of trend patterns.
\end{itemize}

%% file: sections/2-related-work.tex
\section{Related Work}
Prior research relating to search systems in the context of visual data analysis falls under three main themes: (1) general search systems, (2) visual query systems, and (3) NLIs for visual analysis.

\subsection{General Search Systems}
 Broadly, general search systems can be categorized into three types: those built upon structured query languages~\cite{corby:2004,swoogle,Heflin2003SHOEAB,oren:2008}, those that utilize keywords~\cite{falcons,Harth2007SWSEAB,Lei2006SemSearchAS,cimiano:2008,zenz:2009}, and those that are based on natural language processing~\cite{cimiano:2008,Damljanovic2010NaturalLI,fernandez:2008,Lpez2005AquaLogAO,lopez:2006}. Our research extends the capabilities of general search to support intents that involve trends and their quantitative properties, i.e., slope and angle in line charts.

Early research in this area primarily aimed at enhancing conventional text search by integrating metadata using ontological methods to boost both recall and precision~\cite{ciri,Buscaldi2005AWQ,grubar:1993,moldovan:2000}. More recent research has incorporated metadata related to attributes within curated data sources (e.g., synonyms and interrelated concepts) and metadata characterizing pre-existing content, such as visualization techniques, data attributes, and authorship~\cite{olio}. In this work, we crowdsource data on the quantifiable semantics of trend descriptor words, specifically focused on capturing the interplay between language and slope, as well as multi-line trend shapes. This data is leveraged as metadata to boost precision and recall in the context of a search tool for retrieving relevant trend results.

To optimize Q\&A functionality in semantic search, various systems have been developed to precisely identify NL patterns. Common strategies combine statistical methodologies, such as syntactic parsing with semantic processes, to detect ontology-based concepts within user input. For example, QUERIX~\cite{kaufmann:2006} integrates the Stanford CoreNLP parser with WordNet to discern prominent NL phrases in user queries~\cite{klein-manning-2003-accurate}. Other Q\&A frameworks employ linguistic analysis to identify pertinent entities and phrases~\cite{Srihari1999InformationES,panto}. \toolname{} detects analytical trend intents in the search queries and finds trends matching the specified quantifiable properties such as ``sharp decline'' and ``gradual rise'' in univariate line charts.

\subsection{Visual Querying Systems}
Visual querying systems~\cite{lee2019sensemaking} are specifically designed to simplify the process of identifying desired visual patterns within datasets. The ZenVisage~\cite{siddiqui2016zenvisage} visual analytics system was designed for this purpose. Hochheiser \& Shneiderman~\cite{Hochheiser2004DynamicQT} developed a visual query system for time series data, which relied on an interaction method of a ``time box'' by specifying a rectangular region spanning a range of both time and value. Time Lattice supports interactive analysis by customizing a data-cube structure for time-series data with an implicit temporal hierarchy~\cite{Miranda2018TimeLA}. Zhao et al. employed a KD-tree to speed up temporal queries to assist analysts in exploring the local pattern details of interest~\cite{kd-box}. However, these systems do not enable users to employ NL for perusing and exploring the data, nor does it have the capability to understand quantified semantics in trends and patterns. Lee et al. subsequently further explored the space of visual search for data and identified the need for expressive querying and faceted exploration~\cite{lee2019sensemaking}. Our work focuses on NL input as a modality for users to express trend patterns in search, along with faceted browsing to drill up and down the hierarchy of semantic concepts describing these trends.

Continuing this theme of research, Siddiqui et al. introduced ShapeSearch~\cite{siddiqui2021shapesearch} to facilitate the search for specific patterns through sketching, NL, and visual regular expressions. However, the tool does not support the interpretation of quantified semantics for trend descriptors. While basic descriptors like ``up'' or ``down'' are supported, the tool does not accommodate variations in slope and magnitude properties present in trend patterns. In addition, there is no integration of text with the charts to provide additional context to the user during their search task. Our work further explores the nuances of trend patterns and their properties using NL as the modality for expressing such queries. We also integrate text with the search results, along with faceted browsing, to provide additional information and expressivity for navigating the search results.

Bromley and Setlur~\cite{bromley2023difference} recently established an approach for labeling semantic visual features in line charts and proposed its use in supporting the search of shape descriptors for trends. 
We further improve upon this work by carefully designing our experiments and curating our dataset to leverage the precise semantics of trend descriptor words, trend descriptors with modifying adverbs, and multi-line shape trends. In particular, our experiments are aimed at understanding nuances between singleton slope labels (e.g., ``falling''), slope labels with modifiers (``slowly falling''), and multi-line trends that comprise sequential combinations of line segments (e.g., ``peak,'' ``valley''). 

\subsection{NLIs for Visual Data Analysis}
NLIs for visual analysis are designed to facilitate analytical Q\&A~\cite{datatone,thoughtspot,ibmwatson}. They generate charts based on inferred user intent and subsequently introduce ambiguity widgets, allowing users to modify predefined system selections. Both Eviza~\cite{eviza} and Analyza~\cite{analyza} operate upon this premise by integrating contextual inferencing capabilities. Other systems, such as Evizeon~\cite{hoque2017applying} and Orko~\cite{orko}, explore the support of pragmatics within an analytical conversation, leveraging an understanding of the conversational context in play. Flowsense enables NL-based interactions in a dataflow system~\cite{flowsense}. However, the scope of these NLIs focuses on the general support of analytical inquiry and does not consider the interpretation of intents specific to trends and their semantic concepts.

The iGraph system~\cite{ferres2006igraph} focuses on querying trends observed in line graphs; however, the linguistic model employed in the system provides limited support for querying trends and their semantic features pertaining to their quantifiable properties in their slope features. Hoque et al. presented a comprehensive overview of the existing landscape of chart question-answering systems~\cite{hoque2022cqa}. Their survey identified opportunities for supporting more open-ended queries for visual representations and the use of language and semantics to provide more sophisticated models for Q\&A support for data exploration. More recently, the Olio~\cite{olio} hybrid search system combined semantic Q\&A search with document-based exploratory search over data repositories. While the system supports basic analytical intents such as groupings, filters, geospatial, and temporal queries, there is limited support for querying specific semantics for quantifiable concepts such as ``gradual,'' ``sharp,'' or ``plateau'' patterns in trends. Our work further builds upon these search and NLI systems to support the exploration of trends with a comprehensive labeled semantic concept map of trends and their properties.

%% file: sections/3-experiments-data.tex
\section{Creation \& Utilization of Quantified Semantic Label Dataset for Trends} 
To support an expressive analytical experience for exploring relevant trends in a time series dataset, having a quantifiable understanding of the semantics of the trends can be useful. For example, while describing a stock price as ``slumping'' intuitively corresponds to a less severe decline than ``crashing,'' quantifying the nuanced differences between these terms will enable visual data analysis tools to more easily leverage the words' expressive power. 
Bromley and Setlur~\cite{bromley2023difference} previously proposed a crowdsourced dataset of quantifiable visual features for supporting the search of trend shape descriptors. 

However, our experiment design and subsequent analysis extend beyond their work and address shortcomings in several key areas: 
\begin{itemize}
\item \pheading{Greater precision in quantifying slope semantics.} In our experiments, we ask participants to directly label isolated slopes with trend descriptor labels. In Bromley and Setlur's work, participants were instead asked to label line charts (rather than slopes), making it difficult to isolate the direct correspondence between labels and quantifiable slopes. For instance, a participant may have labeled a particular chart with the word ``soaring,'' but each chart contained multiple line segments, making it difficult to ascertain exactly which line segments the participant considered to correspond to a ``soaring'' slope.
\item \pheading{Identification of nuances between trend descriptors.}  We carefully account for and quantify the effect of modifying adjectives or adverbs (e.g., ``fast'') on trend descriptor verbs (e.g., ``falling'') when describing trends. By contrast, Bromley and Setlur treated trend descriptor verbs and modifying adjectives/adverbs as equivalent entities, ignoring the nuances of how these word types can interact to change the semantics of a quantified trend description.
\item \pheading{Support for multi-line shape descriptors.}  We take a thoughtful approach in dealing with multi-line segment shapes in the data (e.g., ``peak'' or ``valley''). In Bromley and Setlur's work, these shape descriptors were treated as equivalent to slope descriptor verbs and associated with a single slope.
\item \pheading{Consideration of semantic relationships between words.} Our analysis of collected trend descriptor word data includes a discussion of suggested semantic relationships (synonym, hyponym, and hypernym relationships) between trend descriptor words. Bromley and Setlur simply treated all words as a flat list with slopes assigned along a continuum.
\end{itemize}

\begin{figure}[hbt!]
    \centering
    \includegraphics[width=\columnwidth]{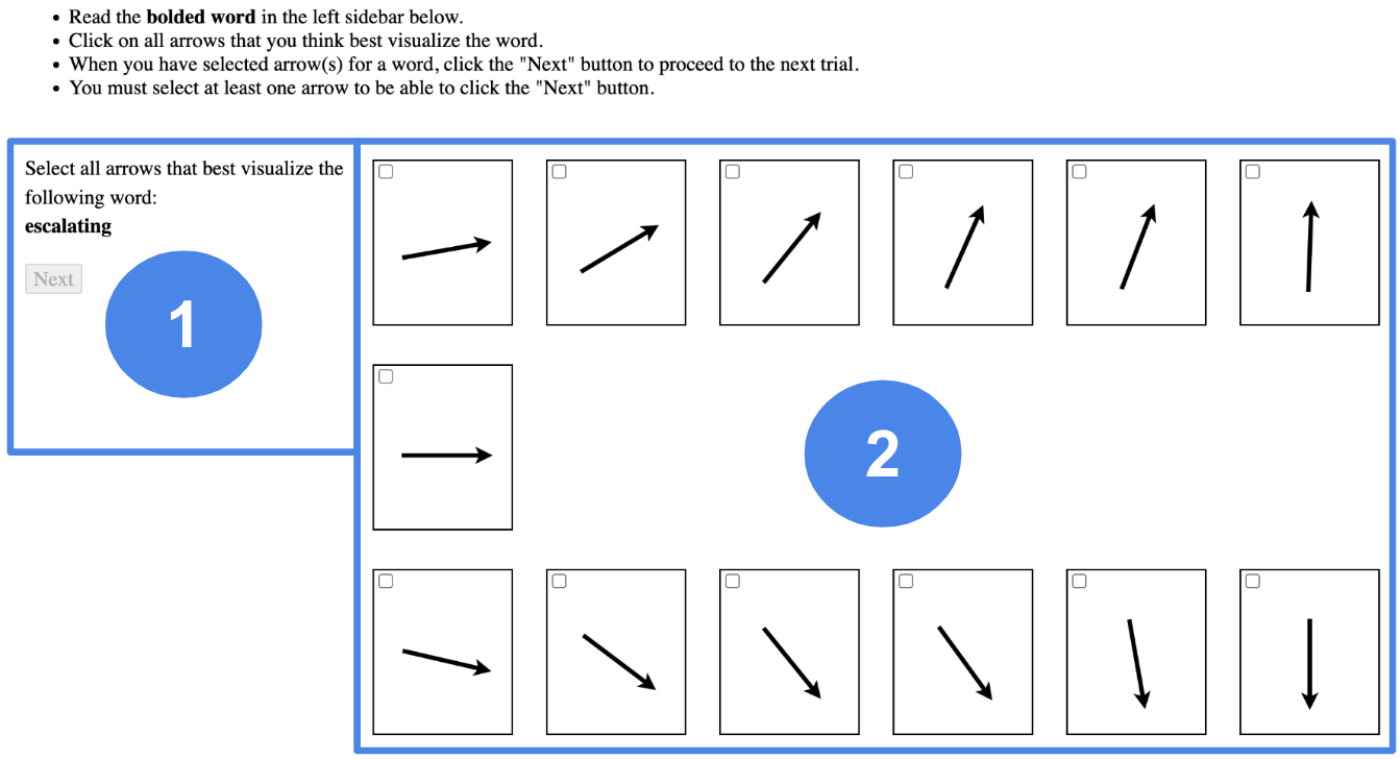}
    \caption{The interface for the data collection web tool used in Experiment 1. (1) The participant is prompted with a word and asked to select all arrows that best visualize the word. Once complete, the participant can click the ``Next'' button to proceed to the next word. (2) The participant is shown $13$ arrows corresponding to an array of angles between -90° and 90°. Clicking anywhere inside an arrow's box applies the current word as a label to the clicked arrow. Note that the interface is similar in Experiment 2, except that in (1), participants are first shown an individual anchor word and then four compound labels, which can be assigned to arrows in turn.}
    \Description[An image of the Experiment 1 interface.]{An image of the Experiment 1 interface. Fully described in text.}
    \label{fig:experiment-1}
\end{figure}

\subsection{Quantified Semantic Label Dataset Collection \& Analysis}
\label{subsec:experiments}
To achieve the advances mentioned above, we designed and conducted three crowdsourced experiments to collect a dataset of quantified semantics for trend descriptor words. We recruited all participants through an internal mailing list at an analytics software company.

We first collected a set of $41$ words used to describe trends that can be mapped to a single quantified slope. We started with the initial list of 21 verbs from Bromley and Setlur's work, then removed 5 words (namely ``accelerating,'' ``intensifying,'' ``decelerating,'' ``subsiding,'' and ``bouncing'') that we deemed could not be mapped to single slope values. We subsequently added six nouns and adjectives from Bromley and Setlur's work that correspond to flat slopes (namely ``flatline,'' ``plateau,'' ``stagnant,'' ``constant,'' ``stable,'' and ``even''). Finally, we augmented this list with 19 additional words sourced from GPT-4~\cite{openai2023gpt} and WordNet~\cite{miller1995wordnet} by querying for synonyms of the words already on the list. The synonyms suggested by GPT-4 were manually inspected by the authors to filter out hallucination responses. Synonyms were only included in the final list if deemed appropriate by all authors.

In the final set, $17$ of these words correspond to negative slopes in time series data (e.g., ``falling,'' ``dropping''), $14$ correspond to positive slopes (``growing,'' ``rising''), and $10$ correspond to relatively flat slopes (see examples above). These words collectively formed the word corpus used for Experiments 1 and 2. For Experiment 3, we included four words from Bromley and Setlur's work that described multi-segment shapes rather than individual slopes (e.g., ``peak,'' ``valley''). We also included 14 additional multi-segment words from GPT-4 with an input prompt, ``\textit{what are the most common multi-segment words similar to `peak' and `valley'?},'' for a total of $18$ such words. As before, these words from GPT-4 were checked for appropriateness by the authors. The data collected from all experiments is included as supplemental material.

\begin{figure*}[htb!]
    \centering
    \includegraphics[width=\textwidth]{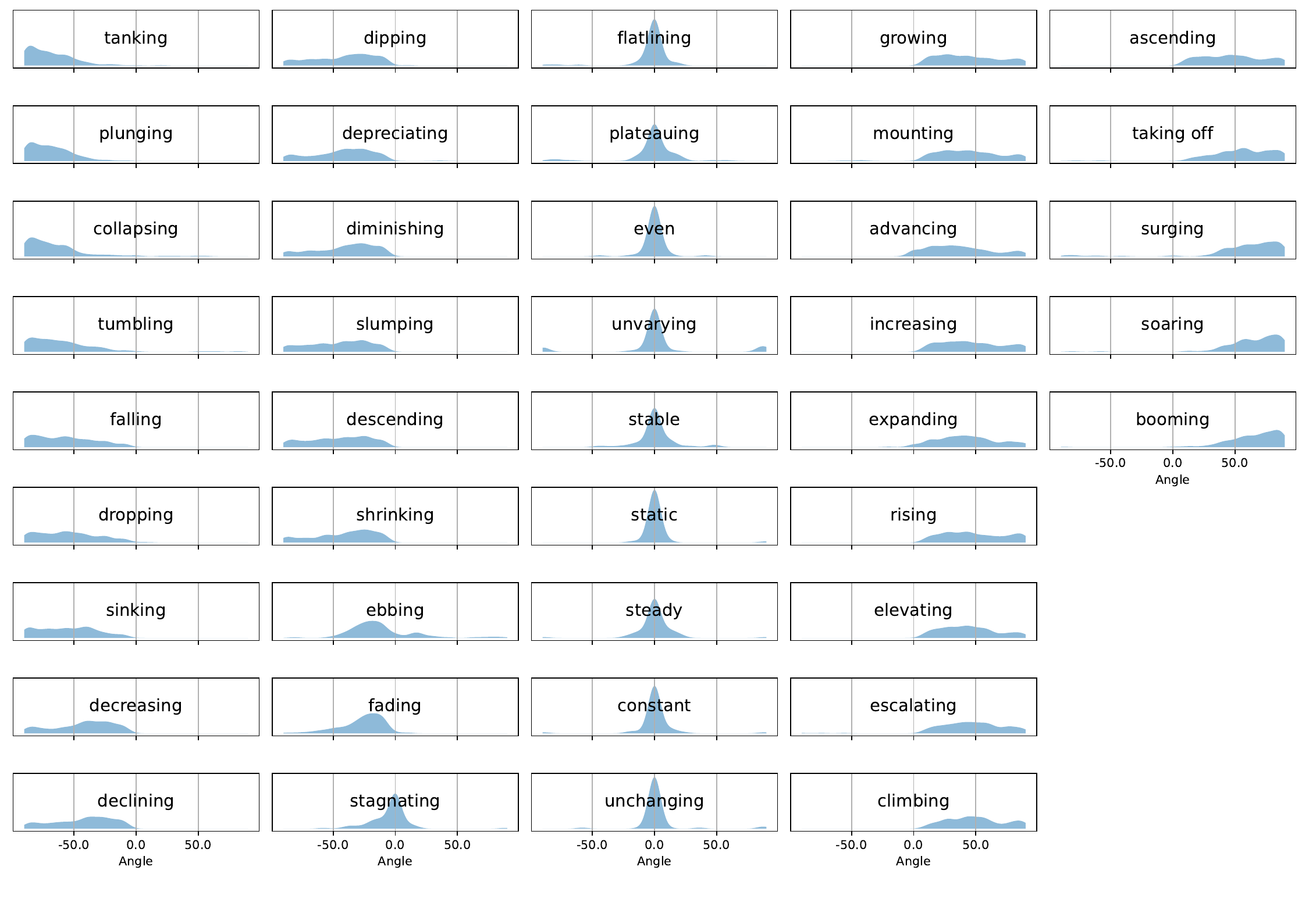}
    \caption{Experiment 1: One-dimensional KDEs indicating probability density for each label over the range of -90° to 90°. Peak probability density was used to sort the labels from the most negative angle (steepest down) to the most positive angle (steepest up) from the top left (``tanking'') to the bottom right (``booming''), respectively.  Note that the distributions are not normal.}
    \Description[An image showing a grid of per-label angle distributions from Experiment 1.]{An image showing a grid of per-label angle distributions from Experiment 1. Fully described in text.}
    \label{fig:exp1_only_label_1d_kde}
\end{figure*}

\subsubsection{Experiment 1: Quantifying Precise Slope Semantics for Individual Trend Descriptors}\hfill

\pheading{Method.} In Experiment 1, our goal was to collect slope information for each of our $41$ single-slope trend descriptor words. The tool's interface (Figure \ref{fig:experiment-1}) presented a set of $13$ arrows whose slopes ranged from 90° to -90° (straight up to straight down, respectively) in increments of 15°. Slopes were jittered by ±7° to provide diversity in the labeled angles. The 7° jitter (just under half of the angle increment between adjacent arrows) was chosen so that the order of the arrows being arranged with increasing magnitude from left to right across the screen would always be preserved. We also made sure that jittered slopes would not exceed the 90° or -90° boundaries since an arrow with such a slope would point backward and thus not make intuitive sense for correspondence with time series data.

The experiment was presented as a sequence of trials, each with one trend descriptor word, rather than as a single page with all $41$ terms present at once. Our reasoning was that we wanted to reduce cognitive burden and avoid visual clutter. For every trial, the participants were shown one of the trend descriptor words (e.g., ``declining'') and were asked to select all of the arrows whose slopes they felt best described that word. The trials were randomized for each participant to mitigate the effects of word presentation order and ensure close to even labeling coverage across words, even if participants did not complete all trials. When a participant clicked an arrow to apply a word label, we recorded the arrow identifier, the arrow slope, the word label, a timestamp of when the annotation occurred, and a unique anonymous participant identifier in a PostgreSQL database~\cite{postgres}. 

\pheading{Analysis.} $80$ participants participated in Experiment 1. Overall, 5,346 angle labels were collected for an average of $67$ labels per participant.  Across the $13$ arrow positions (see Figure \ref{fig:experiment-1}), there was an average of $421$ (min=$301$, max=$531$) labels and $21$ (min=$16$, max=$29$) unique labels per arrow position. The data was analyzed with the goal of estimating a slope distribution for every label. 
We employed Kernel Density Estimation (KDE) \cite{scikit-learn} as a technique for estimating the slope distribution for each label, as KDE is a common tool for estimating the probability density function of a random variable without making assumptions about the nature of the distribution. The Gaussian kernel is a common choice for smoothing KDE data points as the shape is symmetric, it has well-understood mathematical properties, and, notably, the ``bandwidth'' KDE parameter can be interpreted as the Gaussian standard deviation. We selected a bandwidth parameter of 5° to balance between under-fitting and over-fitting the per-label angle data -- a reasonable scale for the range being analyzed (-90° to 90°). Figure \ref{fig:exp1_only_label_1d_kde} shows the slope distribution of each label. It is worth noting that final slope labeling is not particularly sensitive to choices of bandwidth value.  Final labeling is performed by taking the KDE data point with the highest probability density at a given angle (i.e., slope).  Since we use the same bandwidth parameter for all data points, changing the bandwidth would cause all Gaussian distributions to move up and down together, changing the absolute density values but not changing the stack-ranked order. Thus, reasonable changes in bandwidth should not affect the ``winning'' label at any point in KDE space.

\begin{figure*}[htb]
    \centering
    \includegraphics[width=\textwidth]{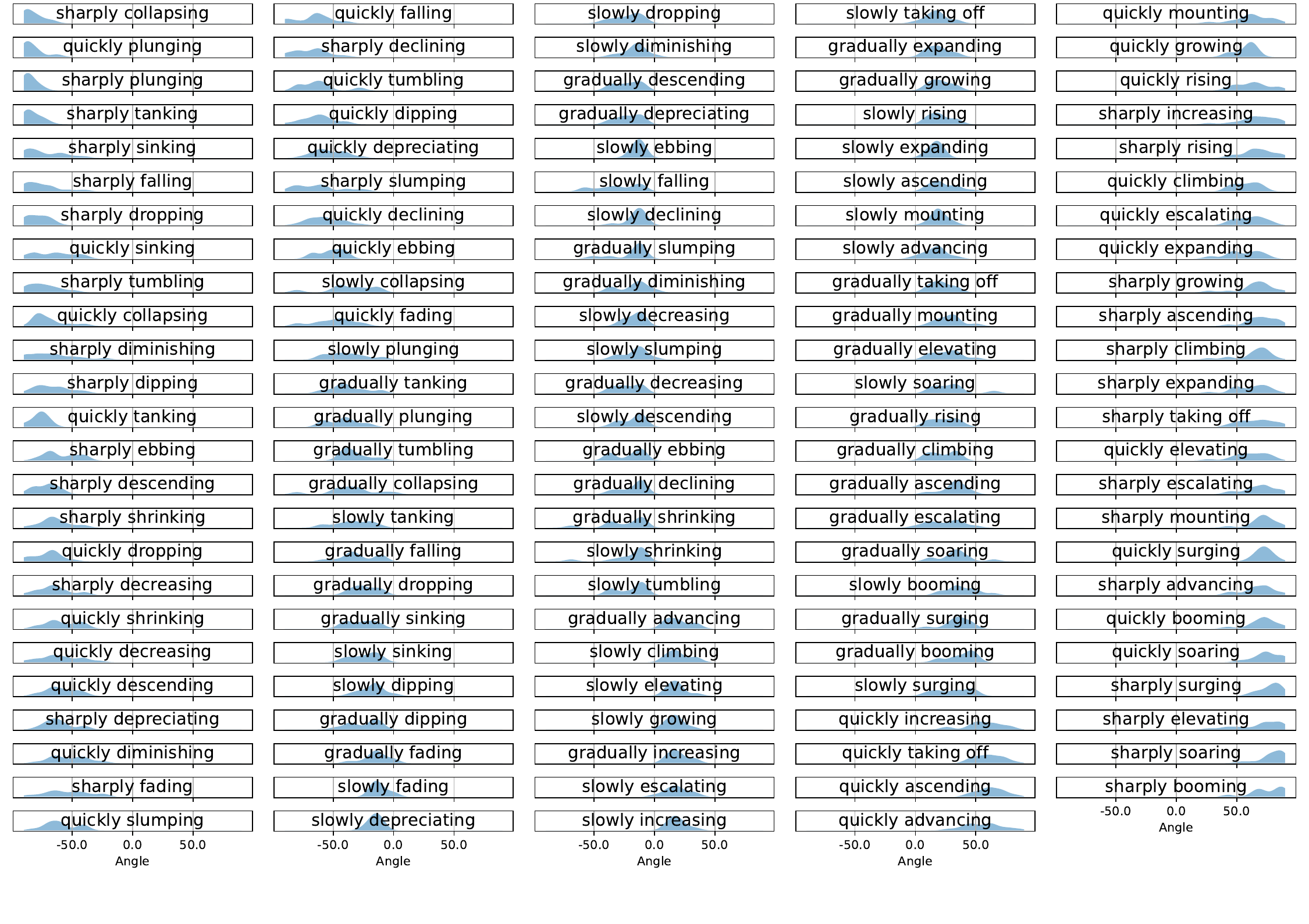}
    \caption{Experiment 2: One-dimensional KDEs indicating probability density for each label over the range of -90° to 90°. Peak probability density was used to sort the labels from the most negative angle to the most positive angle from the top left (``sharply collapsing'') to the bottom right (``sharply booming''), respectively.  Note that the distributions are not normal.}
    \Description[An image showing a grid of per-label angle distributions from Experiment 2.]{An image showing a grid of per-label angle distributions from Experiment 2. Fully described in text.}
    \label{fig:exp2_label_1d_kdes}
\end{figure*}

\subsubsection{Experiment 2: Identifying Nuances Between Trend Descriptors with Modifiers}\hfill

\pheading{Method.} In Experiment 2, our goal was to assess the impact of modifier adverbs on the quantified semantics of two-word trend descriptor phrases, e.g., ``\textit{slowly} falling.'' Of the original set of $41$ words from Experiment 1, we excluded $10$ words corresponding to flat slopes since we would not expect to use modifiers to change a label like ``constant'' or ``even'' to subsequently refer to a different, non-zero slope. We thus only considered the remaining 31 words that corresponded to positive or negative slopes. To create two-word phrases, we selected two adverbs that would make slopes more extreme (``quickly,'' ``sharply'') and two that would make slopes less extreme (``gradually,'' ``slowly'') compared to the anchor label. 

For each trial, participants were first shown a verb and asked to select a single arrow whose slope they felt best matched the word. This single slope was then used as the label ``anchor'' so that participants had a fixed point of reference for the slope labels with modifiers. The participant was then shown four compound labels with the anchor word and the modifiers  (e.g., ``quickly falling,'' ``sharply falling'') and asked to select all arrows that were described by each label. Participants could also choose to discard a compound label if they did not find it to be meaningful (e.g., ``slowly tanking''). Participants were restricted from assigning a compound label to the same slope arrow as the anchor word itself, since we expect the modifying adverbs to always have some effect on the slope of the verbs they modify. In all other cases, a single slope could be assigned multiple compound labels (e.g., a single slope could be both ``quickly tanking'' and ``sharply tanking''). Similar to Experiment 1, the trials were randomized such that anchor words were presented to each participant in a random order. The compound labels with modifiers were listed in a randomized order for each participant (to mitigate bias) but not for each trial to avoid disorienting participants. Whenever a participant clicked an arrow to apply a word label, we recorded the same data as in Experiment 1 with the addition of the current modifier word (or an empty string if the user was setting an anchor arrow).

\pheading{Analysis.} $37$ participants participated in Experiment 2. A total of 2,005 labels (excluding anchor words) were collected. Five participants only picked anchor words; their data was implicitly excluded as there were no modifiers to analyze. The remaining $32$ participants had an average of $62$ (min=1, max=196) labels per participant.  Of the total labels, $144$ were ``trashed'' label modifications that participants did not deem reasonable (e.g., ``\textit{slowly} plunging''). Of the $31$ anchor labels, $28$ labels were trashed for at least one modification (``rising,'' ``sinking,'' and ``climbing'' did not have any trashed modifiers). However, all of these trashed labels were assigned angles by other participants, so there was no general consensus on what label modifications were unreasonable. The remaining 1,861 non-trashed label modifications were spread across all $13$ arrow positions (see Figure \ref{fig:experiment-1}), resulting in an average of $157$ (min=$13$, max=$204$) total label modifications and $33$ (min=$11$, max=$49$) unique label modifications per arrow position. The following rules were used to clean the crowdsourced label data from Experiment 2. In total, $93.4$\% of label/modifier pairs passed these tests and were used for subsequent analysis:
\begin{enumerate}
\item Remove label/modifier pairs where the $\frac{\text{modifier}}{\text{anchor}}$ ratios for ``slowly'' and ``gradually'' were  > $1.0$ or ``quickly'' and ``sharply'' were  < $1.0$; these values were deemed to not be consistent with the common semantic meaning of those words (e.g., the modifier ``\textit{slowly}'' should not make the slope of ``falling'' more extreme).
\item Remove label/modifier pairs where the anchor angle against which the modifiers were compared was 0°; calculating this scalar $\frac{\text{modifier}}{\text{anchor}}$ ratio led to a divide-by-zero error.
\end{enumerate}

Slope analysis of compound labels (e.g., ``\textit{slowly} falling'') was identical to the analysis of singleton labels in Experiment 1. Figure \ref{fig:exp2_label_1d_kdes} shows the KDE distributions of Experiment 2's compound labels.

\begin{figure}[ht]
    \centering
    \includegraphics[width=0.3\textwidth]{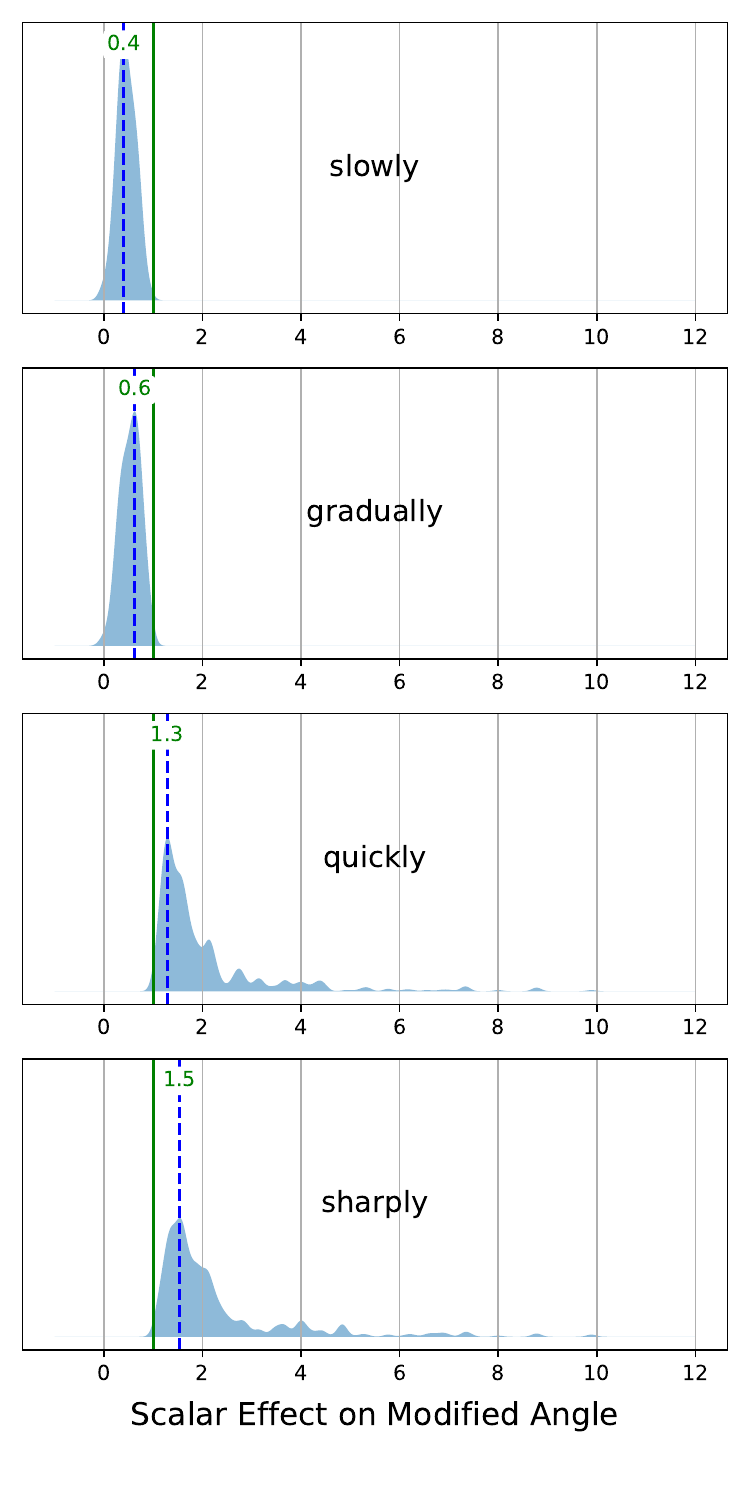}
    \caption{One-dimensional KDEs indicting the scalar range over which different label modifiers scaled the base angle of the label. The solid green line indicates the 1.0 line, and the labeled dotted blue line indicates the scalar value at the peak probability density. Notice that ``slowly'' and ``gradually'' have scalar values between $0$ and $1$ ($0.4$ and $0.6$, respectively), i.e., they reduce the steepness of a label's angle, while ``quickly'' and ``sharply'' have scalar values greater than $1$ ($1.3$ and $1.5$ respectively), i.e., they increase the steepness of a label's angle.}
    \Description[An image showing the distributions of the effect of each label modifier from Experiment 2.]{An image showing the distributions of the effect of each label modifier from Experiment 2. Fully described in text.}
    \label{fig:exp2_modifiers_1d_kde}
\end{figure}

The data from this experiment also enabled the computation of the overall scaling effect each modifier adverb had on each label's associated angle. For example, consider a data point from a single participant indicating that the ``anchor'' angle associated with the label ``dropping'' is approximately -48°.  A subsequent data point from the same participant indicates that the angle associated with the compound label ``\textit{sharply} dropping'' is -88°.  In this case, we can calculate the scalar effect of the modifier ``sharply'' to be -88/-48 = $1.8$. The slope difference between ``dropping'' and ``\textit{sharply} dropping'' can thus be quantified: ``sharply dropping'' is $1.8$ times steeper than simply ``dropping'' for this participant.
We collected these ratios for all angle/modifier data points across all participants and, as before, used KDE analysis to estimate a scalar distribution for every modifier. We again used a Gaussian kernel, and for this analysis, we used a bandwidth parameter of $0.1$ to balance between under-fitting and over-fitting the data.  
As shown in Figure \ref{fig:exp2_modifiers_1d_kde}, in general, ``slowly'' reduces slope steepness by a factor of $0.4$, ``gradually'' reduces slope steepness by a factor of $0.6$, and ``quickly'' and ``sharply'' \textit{increase} slope steepness by factors of $1.3$ and $1.5$, respectively.

\subsubsection{Experiment 3: Supporting Multi-Segment Shapes} \hfill
\label{subsubsec:exp3}

\begin{figure*}[ht]
    \centering
    \includegraphics[width=0.85\textwidth]{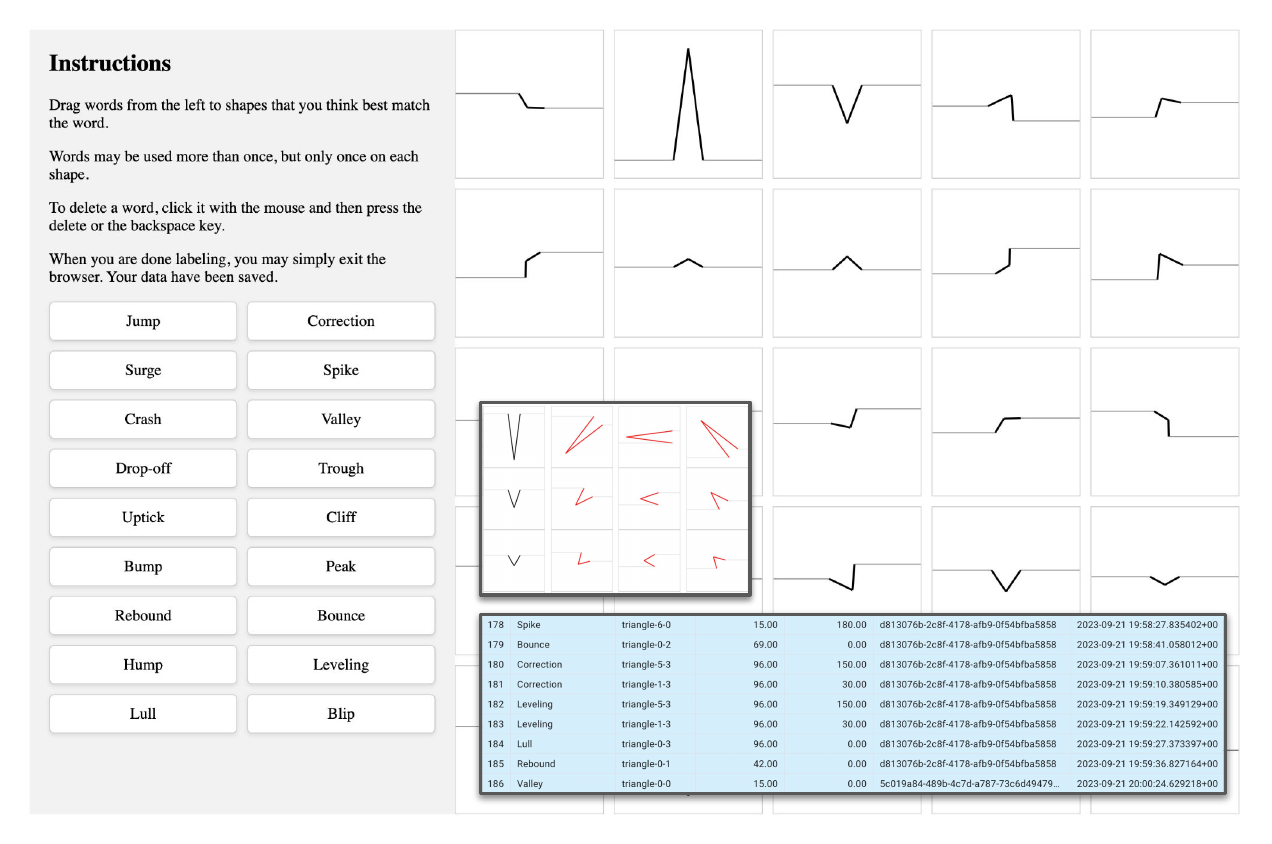}
    \caption{Screenshot of the user interface for Experiment 3. Participants dragged descriptive labels from the left onto shapes on the right.
    Top Inset: Shapes were generated by transforming two-segment angles: angles became more obtuse from top to bottom and were rotated from left to right. Non-monotonic shapes (shown in red) were removed and not shown to the user.
    Bottom Inset: User labels were recorded in a PostgreSQL database; a snapshot of illustrative data rows is shown.}
    \Description[An image of the Experiment 3 interface.]{An image of the Experiment 3 interface. Fully described in text.}
    \label{fig:exp3_ui}
\end{figure*}

\pheading{Method.} In Experiment 3, we aimed to gather labels for different shapes found in time series data. Given that such shapes, in general, can be arbitrarily complex, we focused on performing a relatively thorough sampling of the space of simple shapes consisting of two line segments (e.g., peaks, valleys, plateaus, up-ramps, down-ramps, etc. -- see top inset in Figure \ref{fig:exp3_ui}). We define a \textit{shape} as a pair of connected line segments with varying degrees of (1) inclination angle between the two lines and (2) overall 360° rotation or orientation. Using a similar web interface as before (see Figure \ref{fig:exp3_ui}), participants were asked to label shapes by dragging words from a word list onto the shapes they felt best matched that word. Shape angles spanned the range of 0-180°, and shape rotation spanned the range of 0-360°. Some angle/rotation combinations resulted in non-monotonic shapes; these were removed from the interface (see Figure \ref{fig:exp3_ui}). Participant data was again collected in a PostgreSQL database. Word lists and shapes were randomly arranged to avoid positional bias. 

\begin{figure}[ht]
    \centering
    \includegraphics[width=\columnwidth]{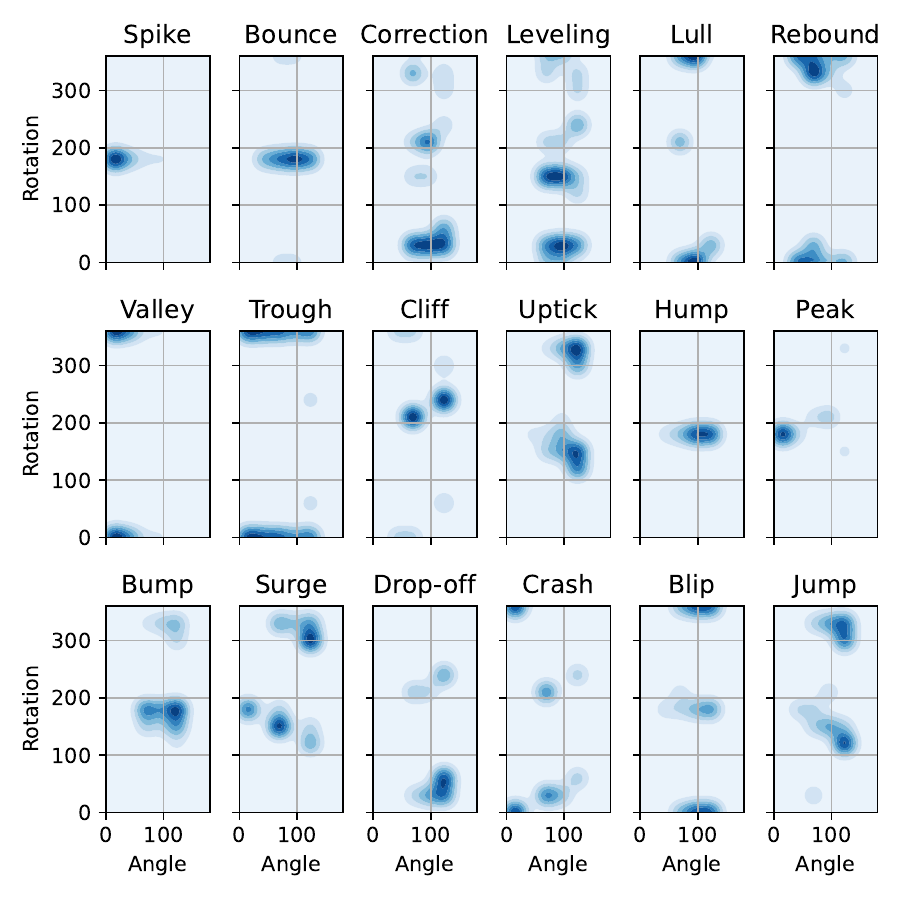}
    \caption{A grid showing a two-dimensional KDE plot for every shape label.  Angles range from 0-180° along the $x$-axis and 0-360° along the $y$-axis.  The $y$-axis is periodic; thus, the probability density is continuous across the 0°/360° border.}
    \Description[A grid of two-dimensional KDEs for shape labels.]{A grid of two-dimensional KDEs for shape labels.  Fully described in the text.}
    \label{fig:exp3_2d_kde}
\end{figure}

\pheading{Analysis.} We collected $347$ labels from $24$ participants for an average of 14 shape labels per participant. The average shape was assigned six different labels  (min=3, max=9), e.g., one shape was assigned the labels ``valley,'' ``trough,'' ``spike,'' and ``crash.'' Conversely, the average label was used to describe eight different shapes (min=$4$, max=$15$), e.g., ``uptick'' was assigned to $12$ different shapes. Similar to Experiments 1 and 2, KDEs were used to describe the label/shape distributions. However, because shapes were parameterized by both angle and rotation, Experiment 3 used two-dimensional KDEs instead of one-dimensional KDEs, resulting in 2D density plots instead of 1D density plots like those from Experiments 1 and 2. We employed Gaussian KDE kernels with a bandwidth of $15$°, an empirically derived balance between under-fitting and over-fitting the data that was a reasonable scale for the values being analyzed (0-360° for rotation and 0-180° for angle).  Also, as discussed above, the final label selection is tolerant to a reasonable range of bandwidth values, so extreme precision is not a significant concern.

Since the shape rotation was periodic, the 2D shape KDEs were made periodic by wrapping them at the 0°/360° boundary. For example, say we wanted to know the data density at a rotation of 3°. To make sure we account for probability density from a point at 355°, we calculated the 3° mark's \textit{virtual} point to be 360 + 3 = 363°, which is influenced by data at the 355° mark after the boundary wrapping.  We then summed the data for the 3° mark and the 363° mark to calculate the final reported value. To calculate periodicity, the 0°/360° boundary overlapped by ±3*bandwidth = ±45°.  The ±3*bandwidth overlap was chosen because three standard deviations (for Gaussian KDE, bandwidth = standard deviation) include $99.7\%$ of the Gaussian distribution. Figure \ref{fig:exp3_2d_kde} shows the 2D KDE plots for Experiment 3.

\pheading{Quantifying Semantic Relationships.}  
We were also interested in exploring how the quantitative labeled data from the experiments could inform the creation of a semantic ontology akin to Wordnet, where semantic concepts are linked by various semantic relations such as synonyms and hypernym/hyponym (superordinate/subordinate) relations~\cite{miller1995wordnet}. Such a structure could be useful for supporting faceted search behavior to drill down or up the semantic hierarchy.

\begin{figure*}[ht]
    \centering
    \includegraphics[width=\textwidth]{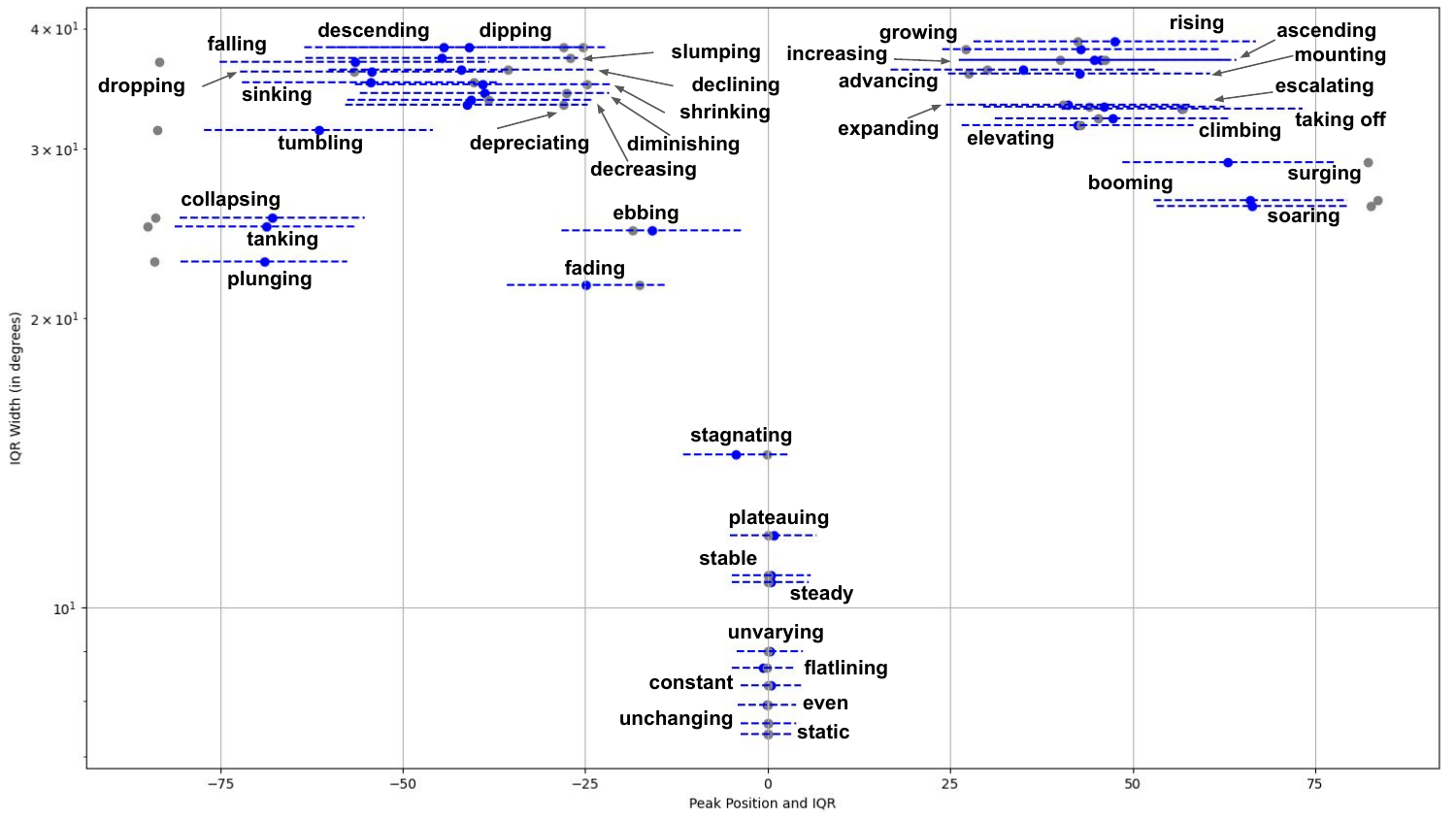}
    \caption{A scatter plot suggesting implicit semantic hierarchies.  The $x$-axis shows the angle range to which the label has been assigned by experiment participants, and the $y$-axis indicates the width of the inter-quantile range (IQR) of the label angle distributions; labels with a broader definition (applicable to more angle ranges) are at the top, labels with narrower definitions are at the bottom.  The blue lines indicate the IQR for each label, the blue dots indicate the center of the IQR, and the gray dots indicate the location of the peak value (mode).  Note that the label distributions are not normal (see Figure \ref{fig:exp1_only_label_1d_kde}), so the mode may lie outside of the IQR.}
    \Description[A scatter plot suggesting implicit semantic hierarchies.]{This figure shows a scatter plot of horizontal bars.  Each bar represents a label from Experiment 1.  The width of the bar is the width of the middle 50\% of the probability density as determined by the Experiment 1 data.  Each bar also shows a dot indicating the middle of the bar and a dot indicating the mode (peak) value. The KDE distributions are not normal; the mode may lie outside the bar.  The bars are vertically arranged such that wider bars are at the top and narrower bars are at the bottom; this allows us to see which angle distributions subsume each other.  At the very top, we see bars for labels such as ``descending'' and ``growing'' and ``slumping''. Below, we see bars for more specific terms like ``plunging'' and ``ebbing'' and ``soaring''. In the bottom middle, we see the terms indicating zero slope, like ``stable'' and ``even''.}
    \label{fig:label_hierarchy}
\end{figure*}

Figure \ref{fig:label_hierarchy} shows a scatterplot of the median, mode, and IQR for the angle distribution of each Experiment 1 label. Wider IQRs are placed higher up, highlighting that the angle ranges of some labels could subsume the angle ranges of other labels, suggesting a hypernym-hyponym relationship (e.g., ``rising'' subsumes ``climbing'' and thus could act as a hypernym).


While the quantitative semantic data shown in Figure \ref{fig:label_hierarchy} suggest synonym or hypo/hypernym relationships that might be derived (\textit{plateauing} >= \textit{stable} >= \textit{steady} >= ...), many of the relationships appear to be more nuanced and suggestive of either partial or multi-category hyponymy and hypernymy.  For example, ``ebbing'' could perhaps be considered a partial hypernym of ``fading;'' there is a partial subsumption relationship, but at the extrema, ``ebbing'' suggests a more positive angle while ``fading'' suggests a more negative angle. Similarly, ``tumbling'' could be seen as a hyponym of \textit{both} ``falling'' and ``diminishing.'' While these statistical methods are quite basic, they suggest that future quantitative semantic analysis could inform the automatic creation of semantic networks and ontologies.


Across the three experiments, we collected 7,554 crowdsourced labels for slopes (5,346 labels), adverb-modified slopes (1,861 labels), and shapes (347 labels), and the data is publicly available in the supplementary material.

\subsection{Labeling Events in Time-Series Data}
\label{subsec:labeling-algorithm}

Once the KDE distributions for both slope labels and shape labels were established, we aimed to use these distributions to label new time series data represented as raw, univariate input signals. 

At a high level, the algorithm for label assignment is as follows: decompose the input signal into linear segments, calculate angles and rotations over those segments, use those angles and rotations to index into the KDEs from the three experiments, and discover appropriate labels. Before we could execute these steps, however, we needed to account for a discrepancy between the aspect ratio of the slope arrows labeled in our data collection experiments and the stock data charts that we plan to use to display stock data in \toolname{}. 

Say that in a chart with a 1:1 aspect ratio, a participant labeled a 45° angle (slope of $1.0$/$1.0$ = $1.0$) as ``ascending.'' If we stretched that same chart to be 100x as wide as it was tall (aspect ratio = 1:100), the participant would probably label that slope as ``flat.'' Conversely if we compressed the chart to be 100x as tall as it was wide (aspect ratio = 100:1), the participant would likely label that slope as ``soaring.''  Thus, it seems reasonable to account for aspect ratio when labeling the perceived ``steepness'' of a line. In this example, we are analyzing a slope where the stock price increased 100\% in 100\% of the time -- that is, it spanned an equal distance along both the $x$ and $y$ axes. In a chart with a 1:1 aspect ratio, this slope would be $1.0$/$1.0$ = $1.0$ = 45°, and we would label it ``ascending'' per the participant's input. However, when we present the chart \textit{visually} to the user, we present it with an aspect ratio of approximately 3:1, \textit{stretching} the $x$-axis such that the same $[0.0,1.0]$ span looks three times longer on the $x$-axis than the $y$-axis. This leads to a \textit{perceived} slope of 1/3 = 0.333 = 18.4°. Thus the user actually sees a line with a slope of 18.4°, not 45°, and the ``ascending'' label looks incorrect because an 18.4° slope, according to our crowdsourced data in Figure \ref{fig:exp1_only_label_1d_kde}, is closer to ``growing'' or ``mounting'' than to ``ascending.'' The core issue is that the participant labeled a \textit{perceived} 45° angle as ``ascending'' and then we showed them a \textit{perceived} 18.4° angle and labeled it (seemingly incorrectly) as ``ascending'' because we did not compensate for the aspect ratio of the visual presentation.  While we realize that there is no ``correct'' data transform as such for this scenario, it seems reasonable that the labeled angle from the input tool should look like the labeled angle in the visual output. Without accounting for the aspect ratio of how the data is presented, the angle and rotation calculations would thus not be true to our collected crowdsourced data.
To resolve this issue, we made two observations:
\begin{enumerate}
\item Participants labeled angles in a user interface with an aspect ratio of 1:1, meaning that the visual space was ``square.''
\item The quality of ``steepness'' is, to a large degree, perceptual and anchored to both the time range we are analyzing and the shape (aspect ratio) of the \textit{chart} and how lines are \textit{drawn} on that chart. Note that this has nothing to do with display size or display device configuration or resolution; this aspect ratio correction is necessary because these labels are \textit{perceptual} labels, not absolute data labels, and as such, we need to correct for the perceived change in angle when a chart is compressed or expanded to an aspect ratio other than 1:1.
\end{enumerate}

Our goal was to design an algorithm that would provide perceptually reasonable results; in particular, slope labels in the output visualizations should visually correspond to the slope labels in the data collection experiments. To this end, we transform the input signal in two ways before we perform the analysis. First, we normalize both the temporal measure ($x$-axis data) and the stock-price measure ($y$-axis data) to span the range $[0.0, 1.0]$, placing both measures on the same scale.

We then scale both measures by the aspect ratio of the expected visual presentation, in this case, 3:1. As a result, the $y$-axis spans the range $[0.0, 1.0]$, and the $x$-axis spans the range $[0.0, 3.0]$. This causes the $[0.0, 1.0]$ span of the $x$ axis to be the \textit{same} perceived distance as the $[0.0, 1.0]$ span of the $y$ axis.  Thus, a line that spans equal distances along the $x-$ and $y$-axes (say, $0.5$ along both) will have a perceived slope of $0.5$/$0.5$ = $1.0$ = 45° and will be labeled ``ascending'' as expected.

\begin{figure}[ht]
    \centering
    \includegraphics[width=0.45\textwidth]{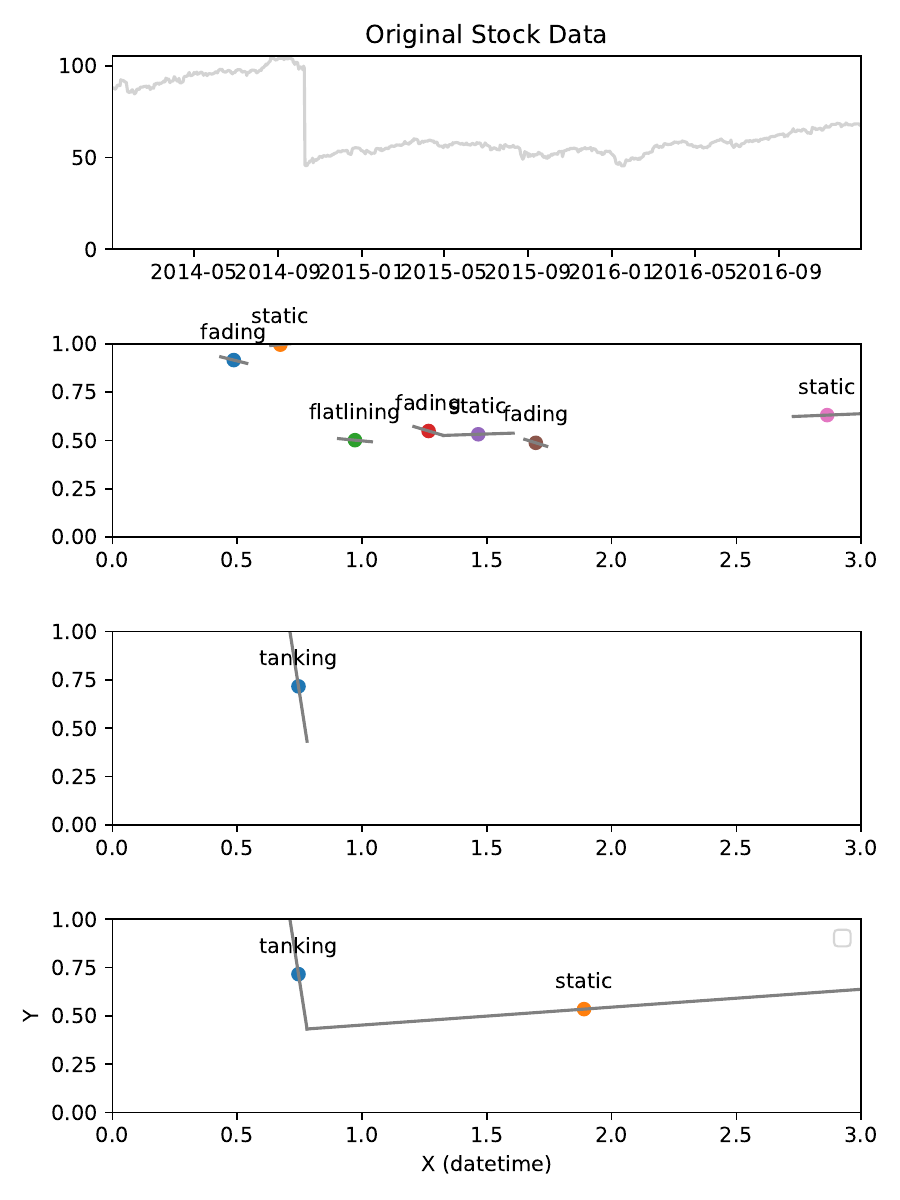}
    \caption{Experiment 1: Angle label assignment. The three sub-charts correspond to the three levels of linearization. The \textit{x} axis indicates time, and the \textit{y} axis indicates stock price.  The original stock data (top) uses the original date and stock value for the \textit{x} and \textit{y} axes, respectively.  The remaining three charts use normalized \textit{x} and \textit{y} values, scaled by the chart's 3:1 aspect ratio, resulting in an \textit{x} range of $[0.0, 3.0]$ and a \textit{y} range of $[0.0, 1.0]$. For clarity, only the top $25\%$ of labels are shown.}
    \Description[Experiment 1 angle label assignments.]{This figure shows the original stock data on top, followed by three charts showing labels for different regions of the stock data.  The three charts use different linearization parameters and thus (potentially) label different time resolutions. The first of the three charts labels several short-term events such as ``fading,'' ``flatline,'' and ``static.'' The middle chart indicates a single longer ``tanking'' event.  The third chart indicates two trends: the ``tanking'' event from the second chart and a very long ``static'' label that lasts for approximately three-quarters of the chart's width.  For clarity, only some of the labels are shown.}
    \label{fig:exp1_only_angle_labeling}
\end{figure}

\begin{figure}[ht]
    \centering
    \includegraphics[width=0.45\textwidth]{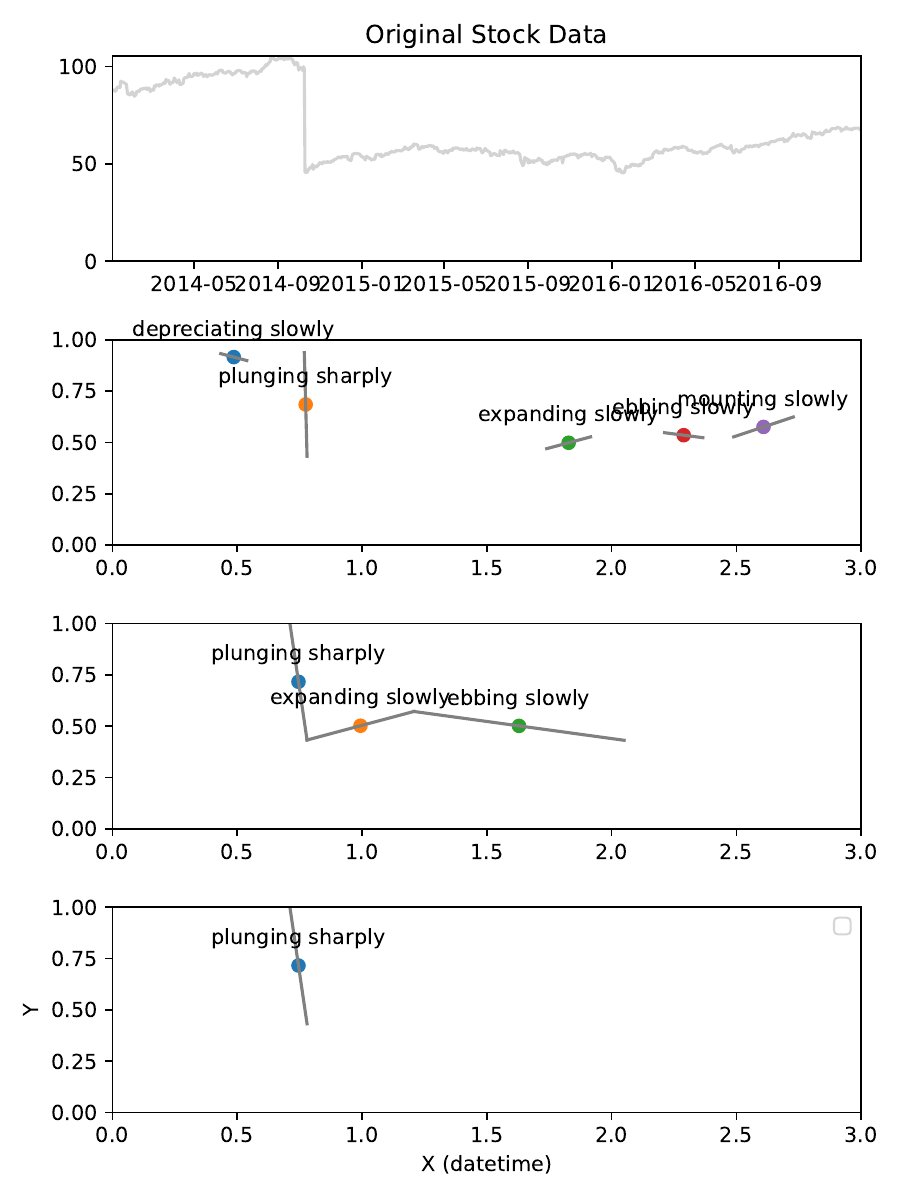}
    \caption{Experiment 2: Angle label assignment.  The three subcharts correspond to the three levels of linearization. The \textit{x} axis indicates time, and the \textit{y} axis indicates stock price.  The original stock data (top) uses the original date and stock value for the \textit{x} and \textit{y} axes, respectively.  The remaining three charts use normalized \textit{x} and \textit{y} values, scaled by the chart's 3:1 aspect ratio, resulting in an \textit{x} range of $[0.0, 3.0]$ and a \textit{y} range of $[0.0, 1.0]$. For clarity, only the top $25\%$ of labels are shown.}
    \Description[Experiment 2: Angle label assignment.]{This figure shows the original stock data on top, followed by three charts showing labels for different regions of the stock data.  The three charts use different linearization parameters and thus (potentially) label different time resolutions. The first of the three charts labels the short-term events as ``plunging sharply,'' ``expanding slowly,'' ``ebbing slowly,'' and ``mounting slowly.'' The middle chart again indicates ``plunging sharply'' and has identified two longer sections for the labels ``expanding slowly'' and ``ebbing slowly''.  The third chart only indicates the ``plunging sharply`` region.  For clarity, only some of the labels are shown.}
    \label{fig:exp2_only_angle_labeling}
\end{figure}

\begin{figure}[ht]
    \centering
    \includegraphics[width=0.45\textwidth]{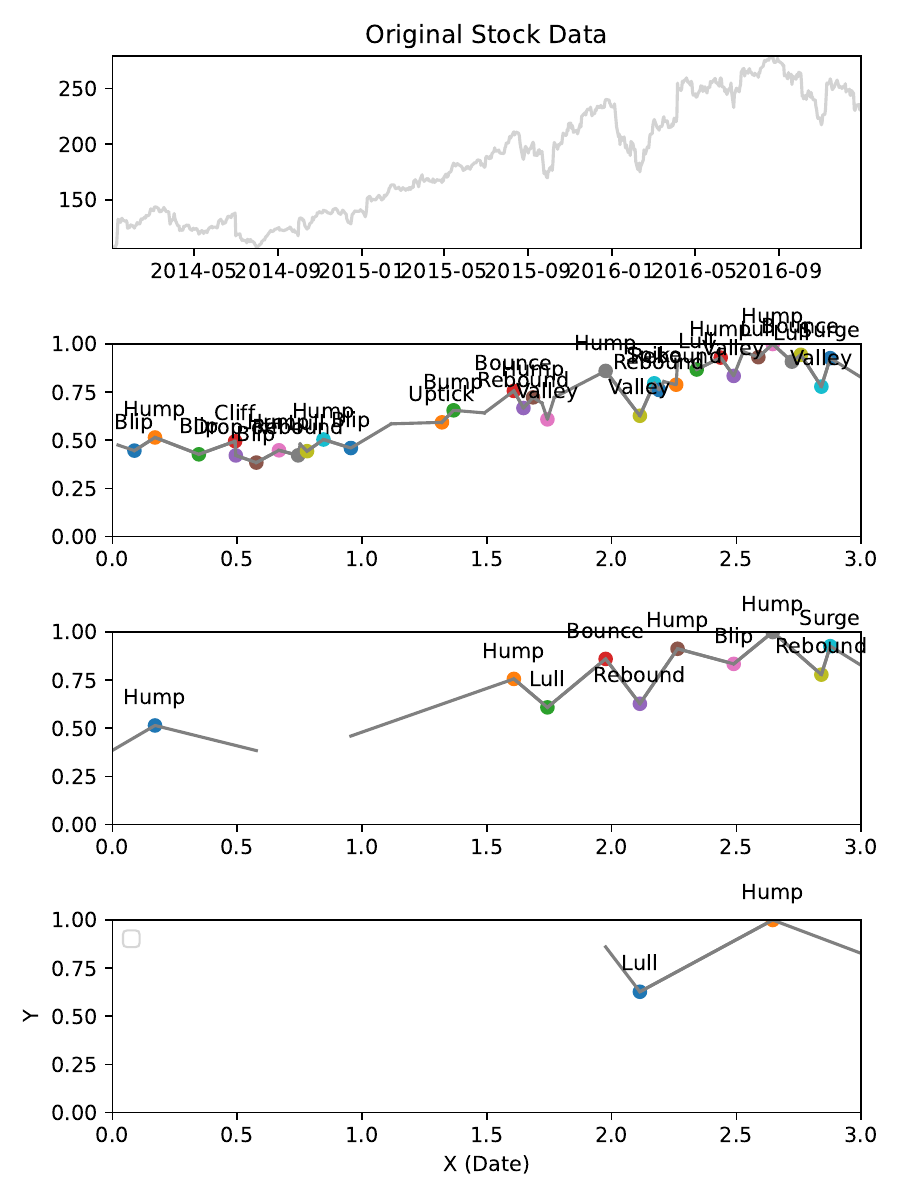}
    \caption{Experiment 3: Shape label assignment. The three subcharts correspond to the three levels of linearization. The \textit{x} axis indicates time, and the \textit{y} axis indicates stock price.  The original stock data (top) uses the original date and stock value for the \textit{x} and \textit{y} axes, respectively.  The remaining three charts use normalized \textit{x} and \textit{y} values, scaled by the chart's 3:1 aspect ratio, resulting in an \textit{x} range of $[0.0, 3.0]$ and a \textit{y} range of $[0.0, 1.0]$. For clarity, only the top $25\%$ of labels are shown.}
    \Description[Experiment 3: Shape label assignment.]{This figure shows the original stock data on top, followed by three charts showing labels for different regions of the stock data.  The three charts use different linearization parameters and thus (potentially) label different time resolutions. The first of the three charts labels multiple short-term events such as ``blip'', ``hump'', ``cliff'', ``uptick'', and ``bounce''.  The middle chart identifies several larger shapes, including multiple ``hump'' labels, multiple ``rebound'' labels, ``lull'', ``bounce'', ``blip'', and ``surge''.  The third chart indicates only two regions: a ``lull and a ``hump' region that overlaps on the second segment of ``lull'' and the first segment of ``hump''. For clarity, only some of the labels are shown.}
    \label{fig:exp3_shape_assignments}
\end{figure}

Following this axis normalization, we proceeded to identify and label temporal stretches of raw input data. The first step was to decompose the input signal into consecutive linear segments using the Ramer-Douglas-Peucker signal linearization algorithm ~\cite{douglas1973algorithms} as implemented in the \textit{rdp} python package (\url{https://pypi.org/project/rdp/}). Epsilon values for the Douglas \textit{et al.} algorithm, which control allowable linearization error and thus the length and scale of the linear stretches, were empirically chosen as $0.03$, $0.1$, and $0.2$ to provide three different linearization resolutions relevant to the stock data under analysis (these values could be adjusted for specific datasets and analyses).

We then determined labels for these linear segments. For single-segment temporal stretches, we first calculated the slope of the segment. We used that slope to index into all the single-label (Experiment 1) or compound-label (Experiment 2) 1D KDEs. The label whose KDE returned the highest probability density was chosen as the label for that segment. Since the KDE models are a continuous surface (1D for Experiments 1 and 2, and 2D for Experiment 3), 
any data point (i.e., line segment) -- even one that is far away from all labels
-- will always return a non-zero probability density score, resulting in \textit{some} (possibly inappropriate) label. To resolve this issue, we took the set of all segment labels (one label per segment), sorted them by their probability density, and only used the top-scoring
75\% of labels for the \toolname{} database. Examples of segment labeling with single-label and compound-label are shown in Figure \ref{fig:exp1_only_angle_labeling} and Figure \ref{fig:exp2_only_angle_labeling}, respectively. For two-segment (shape) temporal stretches, the process was very similar. We first calculated the angle and rotation of each shape. We then used the angle and rotation to index into all the 2D label (Experiment 3) KDEs and kept the top $75\%$ of labels. Examples of shape labeling are shown in Figure \ref{fig:exp3_shape_assignments}.

\begin{figure}
    \centering
    \includegraphics[width=0.85\columnwidth]{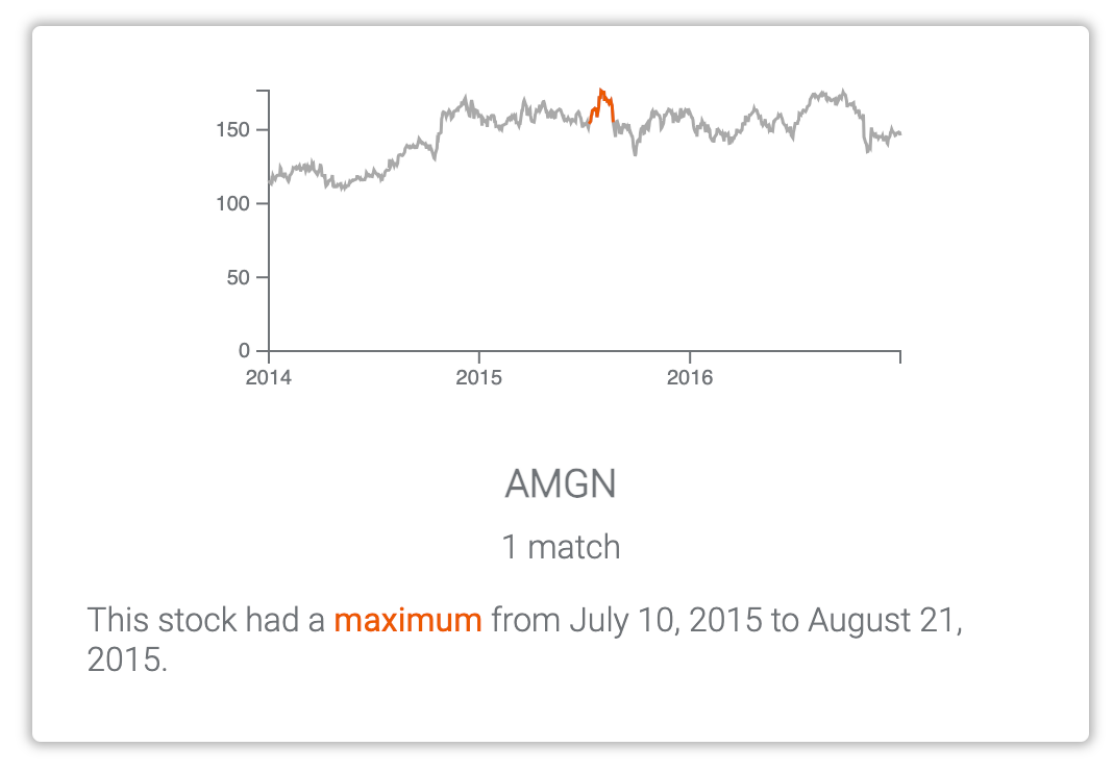}
    \caption{The highlighted event shows the global maximum for this stock over the entire time series.}
    \label{fig:superlatives}
    \Description[An image of a result tile from the SlopeSeeker system interface, showing where a stock has a ``maximum.'']{An image of a result tile from the SlopeSeeker system interface. From top to bottom: (1) A line chart showing a gray line indicating a stock price from the beginning of 2014 through the end of 2016. A small peaked region of the line is colored in red. (2) A stock ticker, ``AMGN.'' (3) The text "1 match". (4) The text "This stock had a maximum from July 10, 2015 to August 21, 2015". The word maximum is in bold and colored in red.}
\end{figure}

To support the querying of superlative features within a trend (e.g., ``maximum,'' ``minimum,'' ``highest point''), we additionally identify the highest and lowest values over the length of the time series data. An event consisting of 15 days before and after the maximum or minimum is subsequently incorporated into the event label to 
result in month-long labeled events that are visually easy to locate (see Figure \ref{fig:superlatives}).

\subsection{Visual Saliency Scoring}
\label{subsec:visual-saliency}
\begin{figure}
    \centering
    \includegraphics[width=0.85\columnwidth]{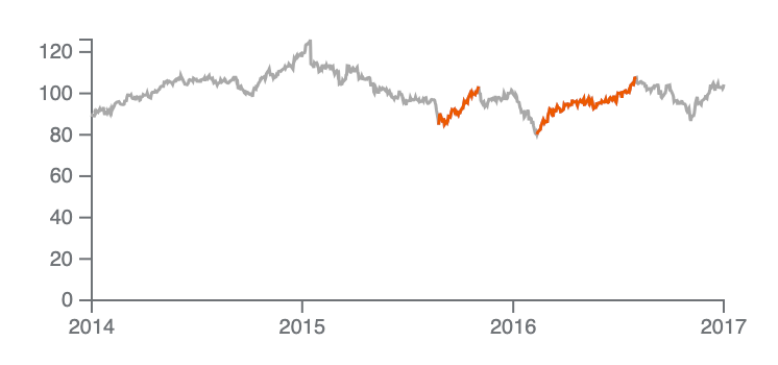}
    \caption{The two highlighted events have identical labels based on their slopes, but the event during the year 2015 is less visually salient than the event during 2016.}
    \label{fig:saliency}
    \Description[An image of a line chart, with two increasing portions in red, the latter being longer.]{An image of a line chart showing a gray line indicating a stock price from the beginning of 2014 through the end of 2016. Two regions of the line are colored in red. The second such region is longer than the first and is thus more visually salient.}
\end{figure}

Prior research has shown that line chart annotations that emphasize the most visually prominent features of the chart are more effective at helping readers glean meaningful takeaways~\cite{kim2021captions}. To operationalize this concept during search, we establish a way to quantify the visual saliency of each labeled trend event. Otherwise, it would be difficult to identify the most relevant results for a given search query when several matching results could have the same labels based on slope. Consider the simple case in Figure \ref{fig:saliency}; while both events match a user query of ``gradually increasing'' based on slope, the event that occurred during 2016 intuitively appears more prominent and impactful than the event in 2015.

The intuition behind our visual saliency scoring approach is to view each trend event as a vector that covers some of the encompassing chart's visual space in both the $x$ direction (i.e., the temporal range of the trend) and in the $y$ direction (i.e., the value range of the trend).

\begin{algorithm}
\caption{Visual saliency computation algorithm}
\label{alg:saliency}
\begin{algorithmic}
\For{each trend result (single-segment slopes):}

    Compute the $x$ vector component as the ratio of the entire time range taken up by the trend.
    
    Compute the $y$ vector component as the ratio of the entire data value range taken up by the trend.
    
    Take these two vector components and use the Pythagorean theorem to compute the L2 norm.
\EndFor
\end{algorithmic}
\end{algorithm}

Formally, we use the following equation (or see Algorithm \ref{alg:saliency} for a procedural representation):
$$
\sqrt{\left(\frac{x_{event_{end}}-x_{event_{start}}}{x_{chart_{max}}-x_{chart_{min}}}\right)^2 + \left(\frac{y_{event_{end}}-y_{event_{start}}}{y_{chart_{max}}-y_{chart_{min}}}\right)^2}
$$
where $x_{event_{start}}$ and $x_{event_{end}}$ are the data values on the $x$ axis at the start and end of the event (and likewise for $y_{event_{start}}$ and $y_{event_{end}}$). Using similar notation, $x_{chart_{max}}$ and $x_{chart_{min}}$ are the maximum and minimum data values on the $x$ axis over the entire time period of the chart (and likewise for $y_{chart_{max}}$ and $y_{chart_{min}}$).

Intuitively, trends described by words like ``tanking'' will mostly be short in $x$, in which case the most visually salient results will have the largest change in $y$. On the other hand, we anticipate that trends described by words like ``flatline'' will have little change in $y$, and so their visual salience will mostly depend on the duration in $x$. However, an intuitive correspondence between trend labels and their visual span in $x$ or $y$
does not need to hold for our scoring to provide a useful quantification of visual saliency. Consider again the two events in Figure \ref{fig:saliency}, both labeled as ``gradually increasing''. The event during 2016 is longer temporally (i.e., $x_{event_{end}}-x_{event_{start}}$ is greater) and spans a larger value range (i.e., $y_{event_{end}}-y_{event_{start}}$ is greater), therefore giving it a higher saliency score.

For multi-segment shapes, we compute the $y$ vector component using the max and min values of $y$ over the duration of the shape event rather than the start and end values. More precisely, we use $y_{event_{max}}-y_{event_{min}}$ in the equation above rather than $y_{event_{end}}-y_{event_{start}}$.

The final labeled stock data loaded into the \toolname{} tool contains 8,353 data points (labeled events) for 100 different stocks over a three-year period (2014 -- 2016).

%% file: sections/4-slopeseeker.tex
\section{\toolname{} Tool}
\label{sec:slopeseeker-tool}

\begin{figure*}[htb!]
    \centering
    \includegraphics[width=\textwidth]{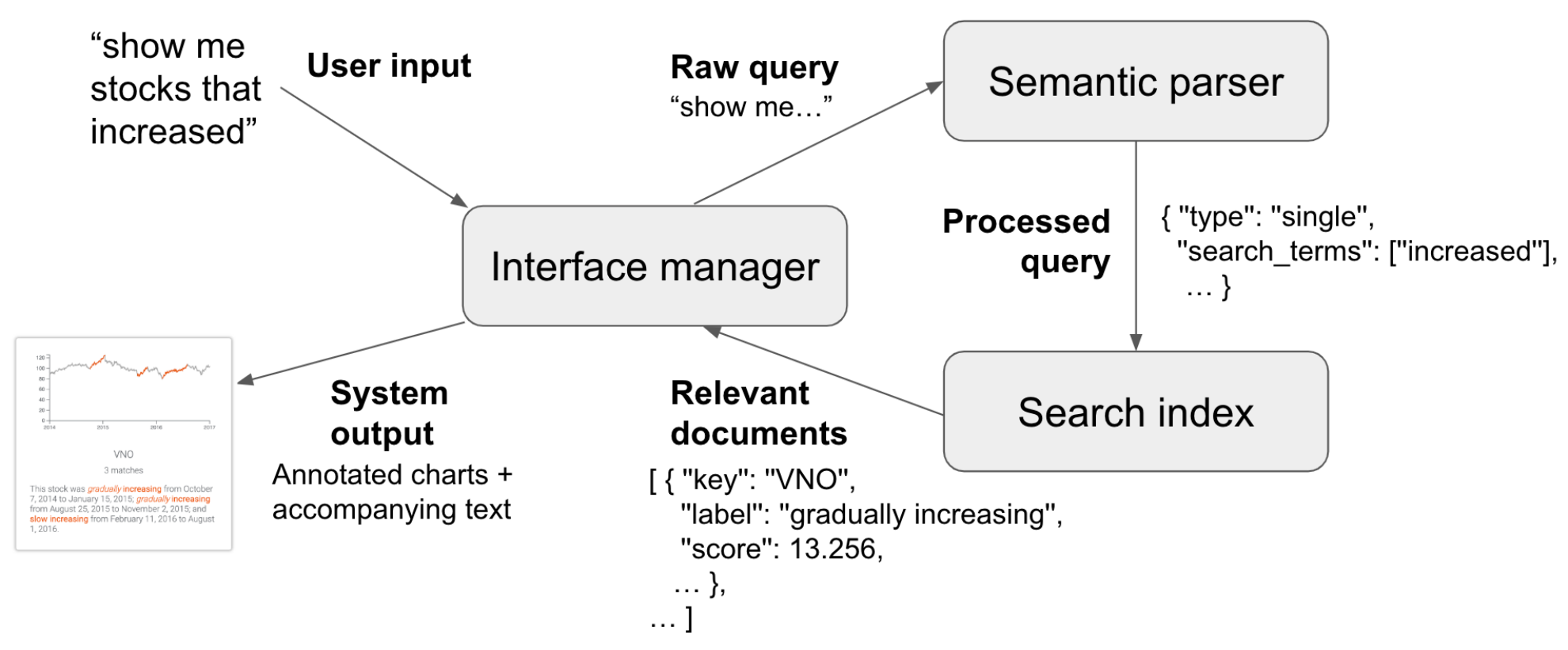}
    \caption{\toolname{} architecture overview.}
    \label{fig:arch}
    \Description[The SlopeSeeker architecture, showing the relationship between the Interface manager, Semantic parser, and Search index.]{A diagram of the SlopeSeeker system architecture. First, user input such as "show me stocks that increased" is taken in by the Interface manager." This raw query is passed to the Semantic parser, which returns a processed query with isolated search terms -- in this case, "increased." The processed query is passed to the Search index, which returns relevant documents (i.e., labeled trend events) such as a particular labeled event where the stock ``VNO'' was "gradually increasing. Finally, the Interface manager takes the relevant documents and uses them to produce the System output of annotated charts and accompanying text.}
\end{figure*}

We developed \toolname{} as a search tool to operationalize our dataset of quantified semantic trend labels. In this section, we first describe the tool's architecture and interface. We then detail the search framework underlying \toolname{}'s functionality and how results are scored. We also outline the different types of trend queries supported by the tool -- single trend events (e.g., ``sharp increase'' or ``peak'') and arbitrary event sequences (e.g., ``up, down, flat'').

\subsection{Architecture Overview}

\toolname{} is implemented as a web-based application using Python and a Flask backend~\cite{flask} connected to a React.js frontend~\cite{react}. For data storage and retrieval, we employ Elasticsearch~\cite{elastic}, a robust distributed search platform built on the open-source Apache Lucene. The platform offers scalability of data and real-time indexing for fast querying. We employ a RESTful API for easy integration with \toolname{}. Figure \ref{fig:arch} illustrates the tool’s architecture, with the following main components: a semantic parser (Section \ref{subsec:semantic-parser}), a search index, and an interface manager that facilitates communication between these back-end components and the front-end interface to implement our end-to-end search framework (Section \ref{subsec:search-framework}).

\subsection{Interface}

\begin{figure*}[htb!]
    \centering
    \includegraphics[width=\textwidth]{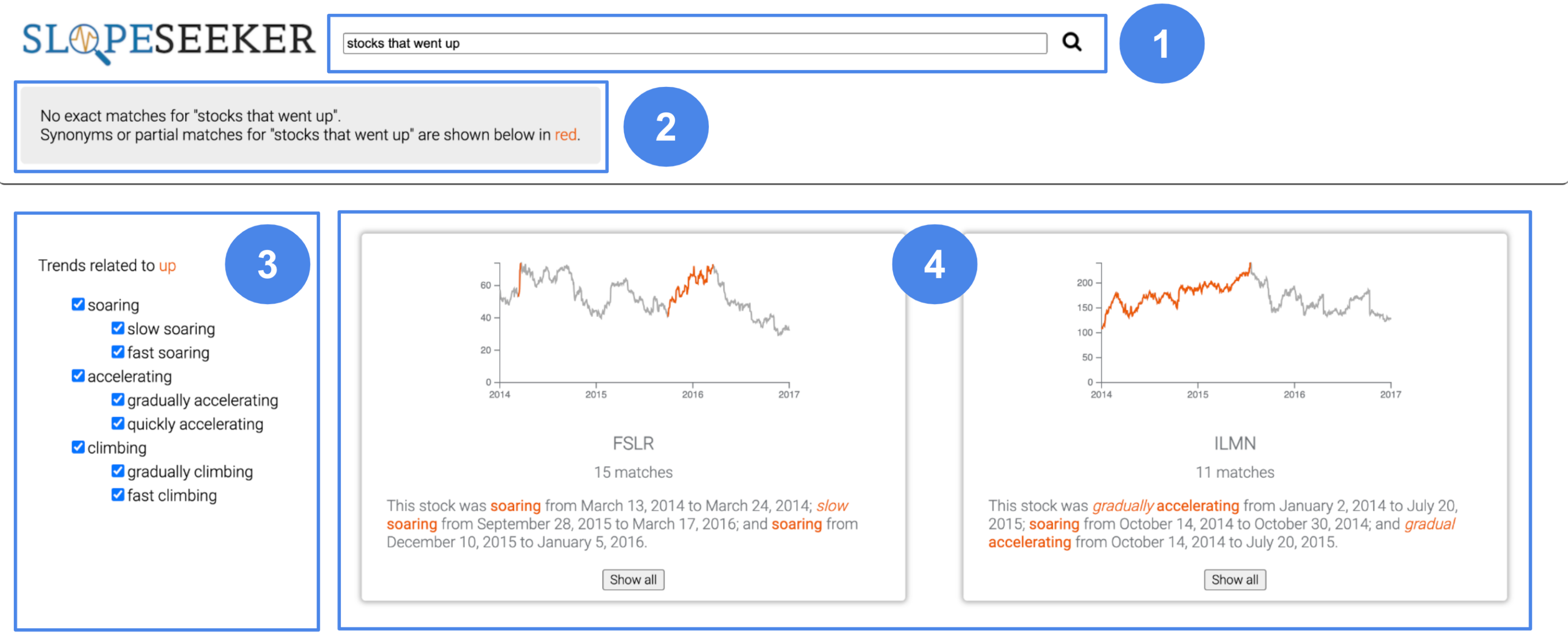}
    \caption{The \toolname{} interface. (1) The search box accepts natural language search queries from the user. (2) The notification box informs the user of inexact matches in the search results. (3) The faceting sidebar enables hierarchical filtering of the results. (4) Search results are shown as interactive tiles containing line charts and textual annotations.}
    \label{fig:interface}
    \Description[An image of the SlopeSeeker system interface.]{An image of the SlopeSeeker system interface. At the top is a search box in which the user has typed ``stocks that went up.'' A gray notification box just below the search bar says, "No exact matches for 'stocks that went up.' Synonyms or partial matches for 'stocks that went up' are shown below in red." Below that and to the left is a sidebar of checkboxes underneath a header that says, "Trends related to up" with the word "up" in red. The checkboxes are nested such that "soaring" is at the outermost layer, with "slow soaring" and "fast soaring" nested underneath. All checkboxes are currently checked. To the right of the sidebar are two result tiles. The left result tile has a line chart with highlighted increasing portions, followed by the text ``FSLR,'' ``15 matches,'' and ``This stock was soaring from March 13, 2014 to March 24, 2014, slow soaring from September 28, 2015 to March 17, 2016, and soaring from December 10, 2015 to January 5, 2016." All instances of the word "soaring" are in red. The right result tile has a line chart with highlighted increasing portions, followed by the text ``ILMN,'' ``11 matches,'' and ``This stock was gradually accelerating from January 2, 2014 to July 20, 2015, soaring from October 14, 2015 to October 30, 2014, and gradual accelerating from October 14, 2014 to July 20, 2015.'' All instances of the words ``soaring'' and ``accelerating'' are in red.}
\end{figure*}

\toolname{}'s interface (Figure \ref{fig:interface}) is designed to provide an experience similar to that of a common web search engine. The user is presented with a search box (Fig. \ref{fig:interface}.1), enabling them to enter a query to search for a trend of interest. After a query has been typed and the Enter key is pressed (or the magnifying glass search button is clicked), results appear as tiles below the search bar (Fig. \ref{fig:interface}.4). Each tile corresponds to one stock and shows a line chart of the stock price over time, the stock ticker, the number of matches for the input query for that stock, and
text snippets (composed into a single sentence)
describing up to the three highest-scoring matches for the stock. The time periods corresponding to these highest-scoring matches are also emphasized in a \textcolor{chart_red}{red} color in the line chart. 
Each emphasized chart segment is interactively and bi-directionally linked with its dedicated text snippet, which describes the corresponding segment's trend label, start date, and end date. Hovering over a chart segment fades out other emphasized segments and will highlight the corresponding text in gray; hovering over a text snippet works similarly. If a stock has more than three matches, the user can expand the tile to show a list of the rest of the matches; hovering over each list item then highlights the corresponding trend in the line chart.

When results do not exactly match the user input, a notification box (Fig. \ref{fig:interface}.2) informs the user which terms are not being matched exactly. The faceting sidebar (Fig. \ref{fig:interface}.3) allows users to optionally filter the results to only include specific labels of interest. The checkbox filters are nested hierarchically by individual label families, e.g., ``soaring'' is a parent of both ``slow soaring'' and ``fast soaring,'' based on the notion of semantic hierarchy discussed at the end of Section \ref{subsubsec:exp3}.

\subsection{Semantic Parser}
\label{subsec:semantic-parser}
We implement a semantic parser module for parsing trend search queries that contain semantic labels, attributes, and temporal filter attributes. Given that the premise of \toolname{} is to demonstrate the utility of the quantified semantic trends dataset in the context of a search tool, we focused on supporting the interpretation of queries specifically intended to search for trends within the stock data. Prior research has demonstrated the effectiveness of a semantic parser in converting NL into a structured representation, which allows for explicit reasoning, reduced ambiguity, and consistent interpretation~\cite{wong-mooney-2006-learning}. Semantic parsers also provide the convenience of better traceability and are performant for structured tasks. Future work could consider combining both semantic parsers for structured tasks and LLMs for open-ended tasks in the context of a more comprehensive analytics tool.

We implement our semantic parser using an open-source Python NLP library, SpaCy~\cite{spacy}, that employs compositional semantics to identify tokens and phrases based on their semantics to create a valid parse tree from the input search query. The parser takes as input the individual tokens in the query and assigns semantic roles to these tokens. The semantic roles are one of four categories: (1) \texttt{event\_type} (single or multi-sequence), (2) \texttt{trend\_terms} (e.g., ``tanking'' or ``plateau''), (3) \texttt{attr} (data attribute names such as stock ticker symbols or company names), and (4) \texttt{date\_range} (absolute and relative data ranges). The tokens and their corresponding semantic roles are translated into a machine-interpretable form that can be processed to retrieve relevant search results in \toolname{}. For an input search query, ``Show me when Alaska Airlines was tanking before November 2016,'' the parser output is as follows:
\vspace{-.8mm}
\begin{minted}[
        linenos=false,
        xleftmargin=10pt,
        tabsize=4]{js}
{     
    'event_type': 'single',  
    'trend_terms': ['tanking'],
    'attr': 'alaska airlines',  
    'date_range': {'lt': '2016-11-01'}  
                  // "before November 2016" 
}
\end{minted}

\subsection{Search Framework}
\label{subsec:search-framework} 

The goal of the search framework is to take the \texttt{trend\_terms} tokens identified by the parser (as well as the \texttt{attr} and \texttt{date\_range}, if applicable) to return relevant results.
Each labeled trend event is considered an Elasticsearch ``document'' in our search context. Documents are the basic units stored in an Elasticsearch index. Once added to the search index, indexed documents can be first retrieved and then ranked according to a match score.
We combine Elasticsearch's built-in scoring logic with our own visual saliency score to produce a scoring mechanism tailored to our use case. Finally, matching documents are grouped by their parent chart for presentation to the user.

\subsubsection{Indexing}
\label{ref:subsubsec-indexing}
The indexing phase creates indices for each document in a dataset along with their metadata. Each of the $n$ documents (i.e., each labeled event -- a portion of a line chart identified by a chart ID, start point, end point, and set of labels) is represented as a document vector $d_i$ where:

$$
\mathcal{D} = \{d_1, d_2, \cdots, d_n\}
$$
We also store sets of string tokens from each document vector to support both partial and exact matches at search time:
$$
\mathcal{S} = \{s_1, s_2, \cdots, s_n\}
$$
\noindent where $s_i = \varepsilon\left(d_i\right)$ for an encoding function $\varepsilon$ that converts each document vector into a set of string tokens. The original vectors $\mathcal{D}$ and encoded tokens $\mathcal{S}$ are stored in the semantic search engine index by specifying the \textit{mapping} of the content, which defines the type and format of the fields in the index. In other words, each semantic trend label and its associated stock data are stored as tokens in the search index in multiple processed formats (i.e., in different fields), enabling fast and flexible retrieval at search time. This indexing enables full-text search on the labels in the index, supporting exact-value search, fuzzy matching to handle typos and spelling variations, and n-grams for multi-word label matching. A scoring algorithm, tokenizers, and filters are specified as part of the search index \textit{settings}. 

\subsubsection{Search (Individual Documents)}
The search phase can be conceptualized as having two steps -- retrieval and ranking. For retrieval, consider a user input query $q$ that is represented as a query vector $\hat{q}$ with query tokens $q_1, q_2, \cdots, q_j$. We encode $\hat{q}$ into string tokens using the same encoding function $\varepsilon$ from indexing, such that $\hat{s} = \varepsilon\left(\hat{q}\right)$. The search retrieval process then returns the most relevant $r$ document vectors $\mathcal{R} = \{d_1, d_2, \cdots, d_r\}$ based on the degree of overlap between the set of query string tokens $\hat{s}$ and the document string tokens in $\mathcal{S}$. Specifically, the scoring function $r_{max}$ maximizes search relevance as follows:
$$
\{d_1, d_2, \cdots, d_r\} = r_{max\ i \in \{1, 2, \cdots, n\}} |\hat{s} \cap s_i|
$$
For search inputs that contain both a noun/verb descriptor (e.g., ``decline'') and a modifying adjective (e.g., ``fast''), we subsequently filter out partially matching documents that contain only the adjective. This logic would prevent a query of ``fast decline'' from returning documents labeled ``fast increase'' as partial matches, for example. More formally, if $\hat{s}$ contains at least one token that matches a noun/verb descriptor in at least one document, then every matching document $d_i$ must contain that descriptor in its set of string tokens $s_i$. However, users may still enter search queries consisting only of an adjective and see documents where that adjective is paired with a variety of noun/verb descriptors.

After retrieval, \toolname{} ranks document results based on two components. The first component is how precisely the search term matches the event labels of the document. Consider a document with a single event label. We utilize a simple scoring scheme where this document's score is the frequency with which the search terms occur in its label, divided by the length of its label, i.e., events with longer labels (e.g., those with modifying adjectives like ``slow'' or ``fast'') will be scored higher than events with shorter labels if and only if the additional tokens accounting for the added length match the search terms. (Note that only \texttt{trend\_terms} parsed search tokens affect scoring, while \texttt{attr} or \texttt{date\_range} tokens are simply used during retrieval to filter results.) 

Consider a document $d_1$ with the label ``slow climbing.'' For a search query of ``slow climbing,'' the score for the document would be $\frac{2}{\sqrt{12}} \approx 0.577$ since it has 2 matching tokens and 12 non-space characters in its label, and thus the label score for $d_1$ for this search would be $0.577$. Now additionally consider document $d_2$ with the label ``climbing'' and a query concerning stocks that are ``climbing.'' The label score for $d_1$ will be $\frac{1}{\sqrt{12}} \approx 0.289$ while the label score for $d_2$ will be $\frac{1}{\sqrt{8}} \approx 0.354$, demonstrating how longer labels with the same number of matching tokens are penalized for being less precise matches.

The second scoring component is the visual saliency score of the document's labeled event (Section \ref{subsec:visual-saliency}), and the final composite score used to rank events in the results is then the product of the Elasticsearch and visual saliency components. The visual saliency component of scoring is most useful when there are a large number of matching results for a user query. Consider a case where the user is interested in ``stocks that increased.'' There could feasibly be very many document results with a label of ``increasing'' which will all have identical (or at least very similar) Elasticsearch scores. However, these results are not likely to all be of equal interest to the user. For instance, a short three-day increase in stock price is probably less interesting, both visually and in terms of the analytical task at hand, compared to a three-month increase during which much more stock value was gained. (Note that these could both have similar slopes and thus identical labels.) The visual saliency scoring component thus serves as a tiebreaker to boost results with greater prominence and relevance over others that share identical labels. 

\subsubsection{Bucketing}
\begin{figure}
    \centering
    \includegraphics[width=0.85\columnwidth]{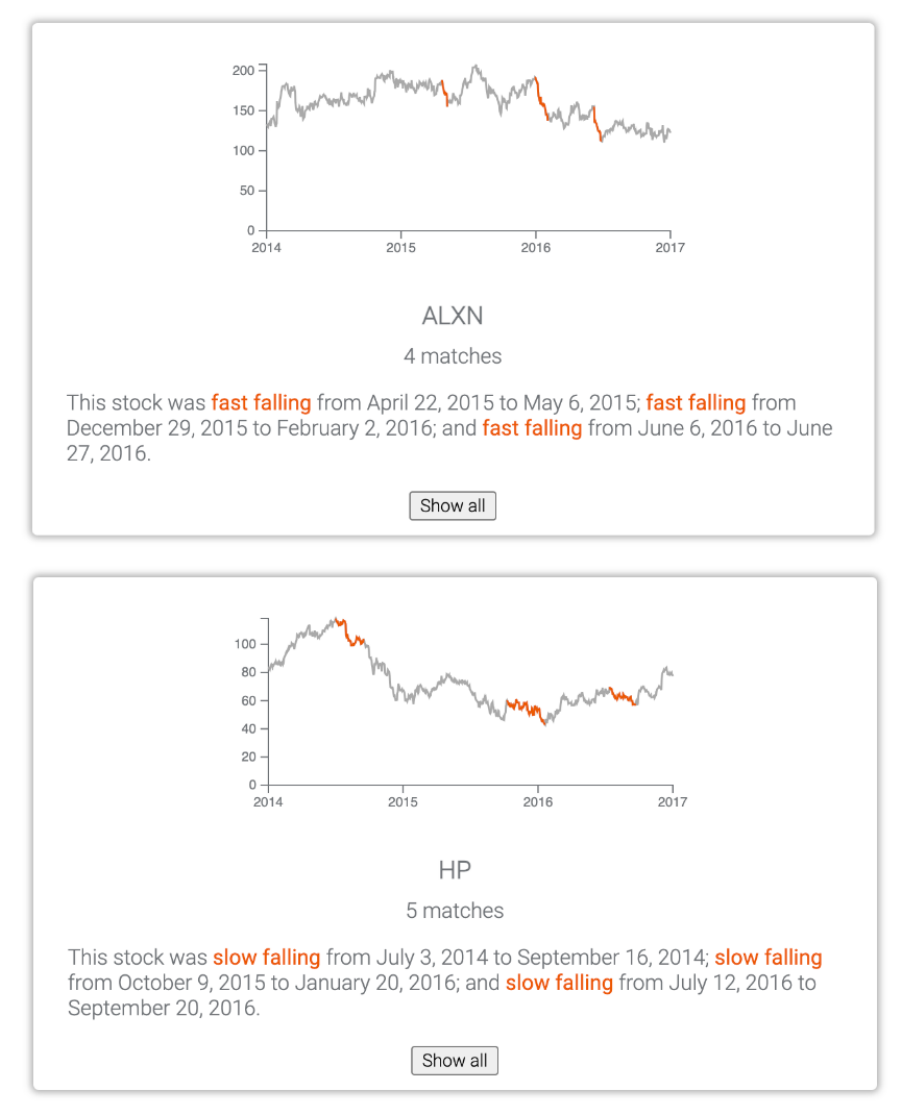}
    \caption{Users can employ modifying adjectives such as ``slowly'' (bottom) or ``fast'' (top) to provide \toolname{} with semantic information about the types of trends they want to see.}
    \label{fig:semantic_resolution}
    \Description[SlopeSeeker results for "fast falling" have steeper declines than those for ``slow falling.'']{An image of two result tiles from the SlopeSeeker system. From top to bottom: (1) A result tile showing matching results for the stock "ALXN" with "4 matches" and the search query ``fast falling.'' The highlighted portions of the line chart have steep downward slopes. It also says, "This stock was fast falling from April 22, 2015 to May 6, 2015, fast falling from December 29, 2015 to February 2, 2016, and fast falling from June 6, 2016 to June 27, 2016." All instances of "fast falling" are in red. (2) A result tile showing matching results for the stock "HP" with "5 matches" and the search query ``slow falling.'' The highlighted portions of the line chart have slight downward slopes. It also says, ``This stock was slow falling from July 3, 2014 to September 16, 2014, slow falling from October 9, 2015 to January 20, 2016, and slow falling from July 12, 2016 to September 20, 2016." All instances of "fast falling" are in red.}
\end{figure}

The indexed data and result scoring are at the level of the document, where each document is an event, i.e., a labeled slope segment. Any individual chart (e.g., stock) could have multiple matching events for a query. 
Events within a bucket are sorted by their composite score. Buckets themselves are also scored; the final score for each bucket is the sum of the composite scores of its individual events, and buckets are presented in sorted order according to this final score. We chose this scheme to create an experience akin to standard document search, where more matches in a bucket bump that bucket higher in the results. Figure \ref{fig:semantic_resolution} demonstrates the differences between high-scoring results (buckets) for the queries ``falling slowly'' and ``falling fast,'' respectively.

\subsubsection{Sequence Queries}
A sequence query consists of a list of trend events (single-word or multi-word) in a specified order, and the sequence query results are generated as follows. First, each individual constituent event is run through Elasticsearch as its own single-word or multi-word query but not yet bucketed. Then, sequences are constructed by taking these results and performing an SQL join based on chart identifier and start/end dates (with a tunable parameter to allow for some temporal delay between adjacent events).

We also have partial matching support for sequences. In particular, we support two types of sub-sequences: edge sub-sequences (e.g., examples for ``up, flat, down'' include ``up'' and ``up, flat'') as well as other in-order sub-sequences (e.g., examples for ``up, flat, down'' include ``flat,'' ``down,'' and ``flat, down'').

We additionally define a scoring scheme for sequences and partial sequence matches. At first, each sequence's score is assigned to be the sum of the composite scores of its constituent segments. Although partial matching sequences have fewer constituent components and will generally have lower composite scores than full matches, we found it beneficial to additionally down-weight the composite scores of partial sequence matches. In particular, we use the following formula where the un-penalized score is notated $\text{score}_0$, the number of events in the sequence being scored is notated $l_{seq}$, the number of events in the query is notated $l_{\hat{q}}$, and the sequence offset is the number of sequential events missing from the beginning of the sequence compared to the query:

$$
\text{score}_0\left(\frac{l_{seq}}{l_{\hat{q}} + \text{offset}_{seq}}\right)^2
$$

\vspace{2mm}
This custom scoring scheme applies two different penalties. First, longer sub-sequences ($l_{seq} \approx l_{\hat{q}}$) are penalized less and thus scored higher than shorter ones ($l_{seq} < l_{\hat{q}}$). For example, if the user is interested in stocks with the pattern ``up, flat, down'' which has length three, a sub-sequence of length two (e.g., ``up, flat'') will be scored higher than a sub-sequence of length one (e.g., ``up'') because the sub-sequence matches more constituent events of the sequence. Second, a non-edge sub-sequence (with a large offset) will be penalized and scored lower than an edge sub-sequence (with zero offset). Continuing with the same example, the sub-sequence ``up, flat'' has zero offset because it begins at the same place as the initial query pattern, but ``flat, down'' has an offset of one since it starts one event later in the sequence. Intuitively, sub-sequence partial matches that begin similarly to the desired sequence from the query should be scored higher than those that end similarly to the desired sequence. Finally, after applying any score penalties to the sequence results and any partial results, all results are bucketed, and the bucket scores are computed as before.

\begin{figure}
    \centering
    \includegraphics[width=0.85\columnwidth]{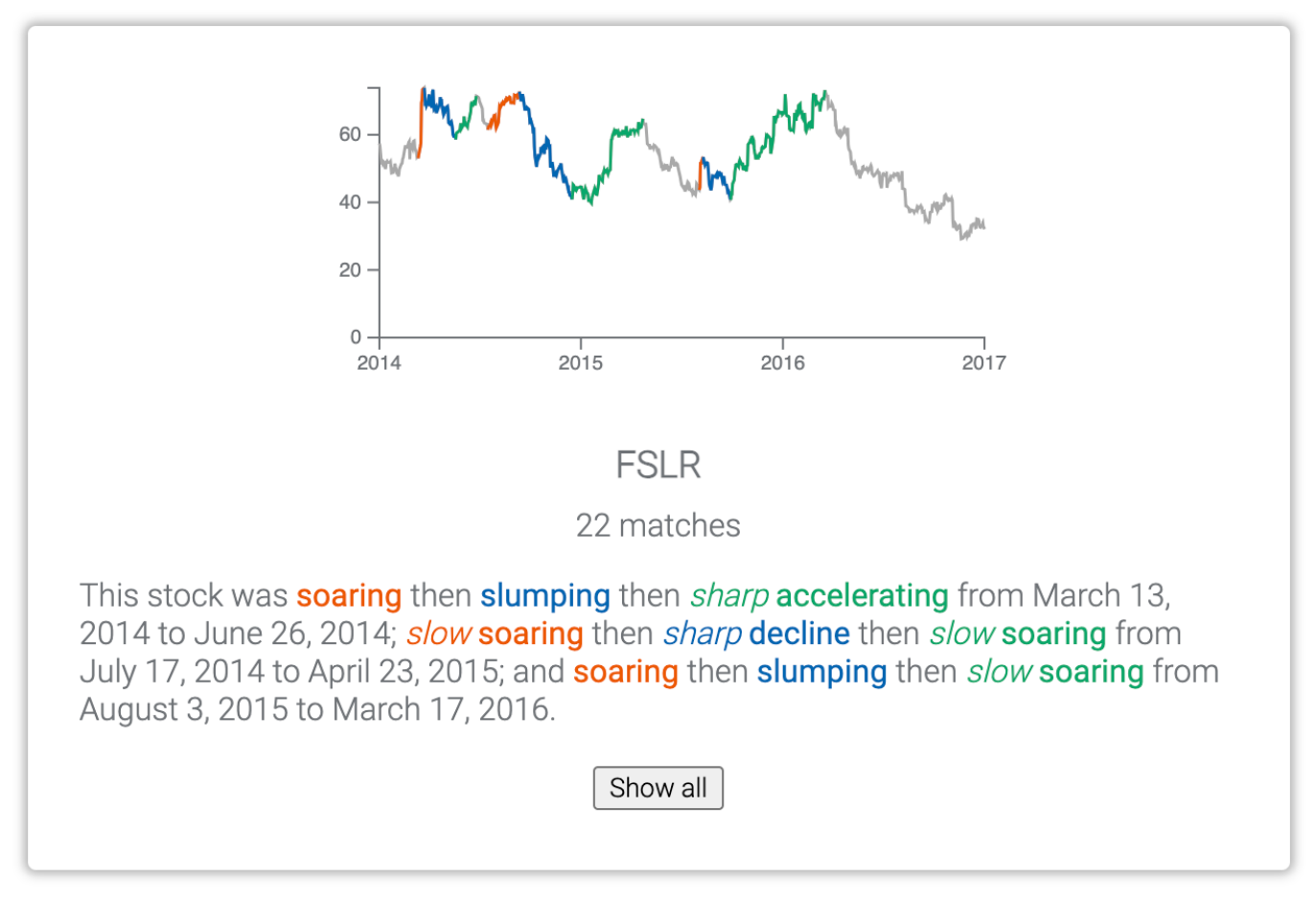}
    \caption{A top search result when the user searches for the sequence ``\textcolor{chart_red}{up}, \textcolor{chart_blue}{down}, \textcolor{chart_green}{up}.'' Each color corresponds to a unique position in the sequence. 
    }
    \Description[A result for the sequence query "up, down, up". Stock portions are in different colors for each of the three sequence parts.]{An image of a result tile from the SlopeSeeker system corresponding to the sequence query "up, down, up." On both the line chart and corresponding text, portions matching the first "up" are in red, portions matching "down" are in blue, and portions matching the second "up" are in green. The tile has the stock ticker "FSLR", has "22 matches" and also says the following. This stock was soaring then slumping, then sharply accelerating from March 13, 2014 to June 24, 2014; slow soaring, then a sharp decline, then slow soaring from July 17, 2014 to April 23, 2015; and soaring, then slumping, then slow soaring from August 3, 2015 to March 17, 2016.}
    \label{fig:sequence}
\end{figure}

%% file: sections/5-evaluation.tex
\section{Preliminary Evaluation of \toolname{}}
Using \toolname{} as a design probe, we conducted a preliminary evaluation to gather feedback on the utility of the quantified semantic label dataset in the context of a search tool for trends.

\subsection{Participants and Setup}
We recruited $12$ participants (P1-P12, six male and six female) through a mailing list at an analytics software company. Participants volunteered their time on a first-come, first-served basis, and due to company policy, they were not compensated for their participation. Based on self-reporting, participants comprised three data scientists, three sales consultants, two product managers, one program manager, one account executive, one UX researcher, and one HR analyst. Half of the participants reported that they perform data analysis on a regular basis (daily or almost daily), while the other half reported that they occasionally perform data analysis (weekly or biweekly). All participants reported that they regularly use a search tool like Google.

All sessions were conducted in person. The \toolname{} tool was hosted on a local server running on the experimenter's laptop\footnote{2.4 GHz MacBook Pro running macOS Ventura 13.5 set to a resolution of 3072 $\times$ 1920.}. The audio, video, and on-screen interactions were recorded for all sessions after receiving permission from each participant.

\subsection{Procedure}
Study sessions lasted about 45 minutes and followed the protocol outline below:

\noindent[$\sim$10 min.]: Participants were given an overview of the evaluation and were asked to self-report their relevant background information in visual analysis. Participants were then briefly introduced to the \toolname{} interface. The introduction included the capabilities of the search tool without providing any explicit NL queries to avoid biasing participants. 

\noindent[$\sim$25 min.]: Participants were given a set of four tasks involving a dataset of labeled time series stock data and were asked to complete the tasks using \toolname{}. The first three tasks were designed to prompt users to utilize at least one of the types of supported queries, i.e., single slopes (downward and upward slopes) and multi-segment shapes. The task prompts were as follows: (1) ``find an instance where a stock gained a lot of value (in a certain year or time frame),'' (2) ``Find an instance where a different stock lost only a small amount of value,'' and (3) ``Find two stocks whose price followed this pattern'' (showing a visual of a valley). The fourth task was open-ended, wherein participants were prompted to identify a stock they would want to invest in based on the historical patterns of stock prices in the dataset. 

\begin{figure}[t!]
    \includegraphics[width=.4\textwidth,keepaspectratio]{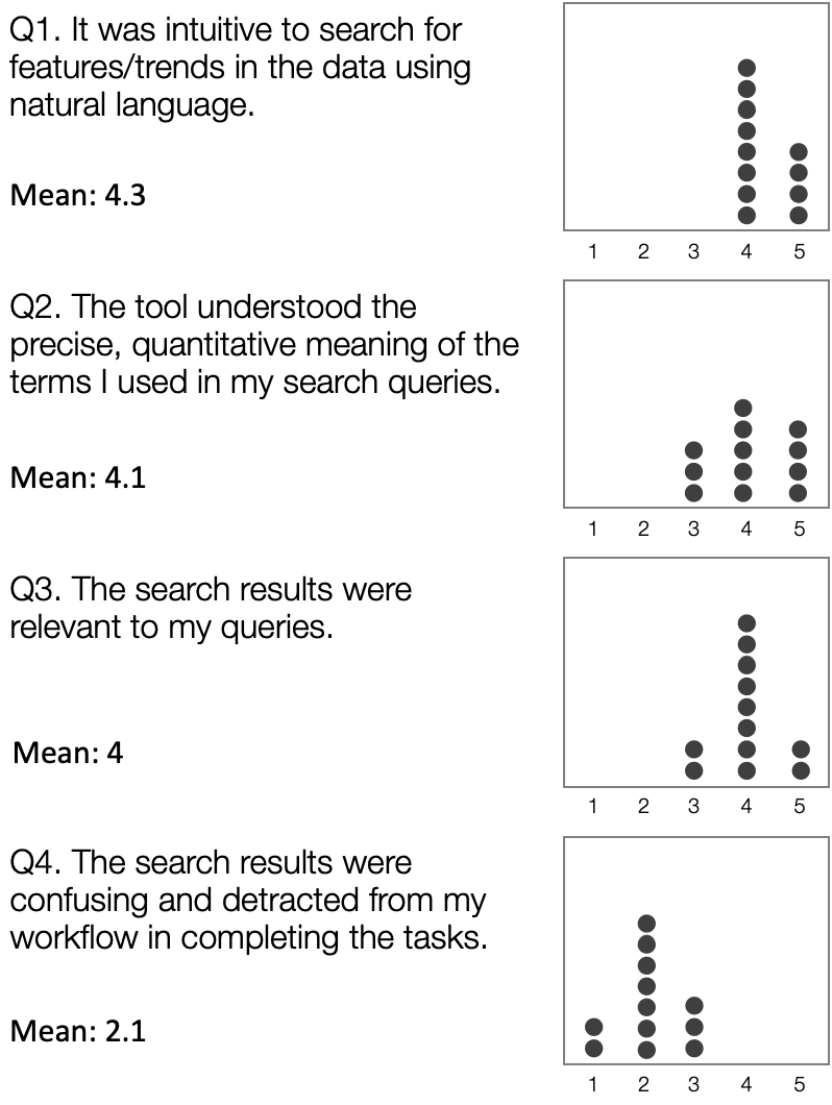}
    \caption{Participant responses to post-session questions about utterance recommendations in \toolname{}. Statements were rated on a scale of 1 (Strongly Disagree) to 5 (Strongly Agree).}
    \Description[Participant responses to post-session questions, generally showing that the tool was intuitive and gave relevant results.]{A chart showing participant responses to four post-session questions. "Q1. It was intuitive to search for features/trends in the data using natural language. Mean: 4.3." To the right is an image showing eight ratings of 4 and four ratings of 5 for Q1, indicating somewhat strong agreement. "Q2. The tool understood the precise, quantitative meaning of the terms I used in my search queries. Mean: 4.1." To the right is an image showing three ratings of 3, five ratings of 4, and four ratings of 5 for Q2, indicating somewhat strong agreement. "Q3. The search results were relevant to my queries. Mean: 4." To the right is an image showing two ratings of 3, eight ratings of 4, and two ratings of 5 for Q3, indicating somewhat strong agreement. "Q4. The search results were confusing and detracted from my workflow in completing the tasks. Mean: 2.1." To the right is an image showing two ratings of 1, seven ratings of 2, and three ratings of 3 for Q4, indicating somewhat strong disagreement.}
    \label{fig:survey_results}
\end{figure}

\noindent [$\sim$10 min.]: The sessions concluded with a post-session questionnaire (Figure~\ref{fig:survey_results}), ten questions from the standard System Usability Scale (SUS) questionnaire~\cite{sus} to help evaluate the prototype's usability, and a semi-structured interview discussing participants' overall experience using \toolname{} for trend search and areas for further improvement. 

 The \toolname{} tutorial, study protocol, and questionnaire are included in the supplementary material. 

\subsection{Results and Discussion}
Overall, participants found \toolname{} to be a useful tool for searching and exploring trends in the stock data. We detail participant feedback and usage behavior with respect to the tasks and open-ended exploration during the evaluation. All participants were able to complete the four tasks, spending between 16--22 minutes (mean: 18 min.). Participants gave \toolname{}~an average SUS score of $79.4$ (a score of $\ge 68$ is considered as an indicator of good usability~\cite{sus}). 

\subsubsection{Intuitiveness of Interface to Search for Trends}
Participants generally agreed that \toolname{} was intuitive for trend exploration (Figure~\ref{fig:survey_results}, Q1). For instance, P2 found the interface and the faceted search to be similar to that of familiar search systems -- ``\textit{This looks similar to Google and Amazon, and I could get started right away.}'' Participants predominantly used the facet filters to see more general and specific trends such as navigating between ``\textit{slight} increase'' and ``increase.'' P7 noted the ``\textit{ability to drill down into a specific trend description}'' by using the faceted search options. Participants also found the display of results in the interface to be useful. P1 commented, ``\textit{I like the tile-like view of all the trends. Helps with quick glanceability, and I know what's going on.}''

\subsubsection{Interpretation of Search Queries}
Participants indicated that \toolname{} appropriately interpreted their search queries for single slopes and multi-segment shapes (Figure~\ref{fig:survey_results}, Q2). However, we noticed that the quality of search results deteriorated when queries included additional information such as relative time periods (e.g., ``did the apple stock go up \textit{recently}?''), subjective concepts (e.g., ``\textit{best} stocks to buy in 2014''), or attributes that were not included in our stock dataset. However, all participants appreciated the text accompanying each trend result that showed the quantifiable trend with the corresponding trend segment in the chart upon hover  (Figure~\ref{fig:interface}.4), commenting, ``\textit{this is pretty neat}'' [P5] and  ``\textit{I can compare the labels for what `cliff' means and see it visually. That's helpful to know what's going on}'' [P11].

\subsubsection{Relevance of Search Results}
\label{sec:results-feedback-relevance}
Generally, participants were in agreement that the search results were relevant to their input queries (Figure~\ref{fig:survey_results}, Q3 and Q4). Participants specifically appreciated that the tool could differentiate between nuances in quantifiable semantics in the trends. P7 cited an example from her session and said, ``\textit{Wow, I asked for trends that hit a cliff in 2014, and it's pretty cool to see an array of results with cliffs in them.}'' Others found \toolname{}'s capability to detect a sequence of trends to be useful. P12 stated, ``\textit{I was curious to see what stocks went up and then suddenly down, and I was impressed that it recognized `suddenly' for `down.'}'' However, there were also some mixed reactions on the relevance of the results, where participants
indicated limitations in the capabilities of the tool, as described in the takeaways below:

\pheading{Consider pragmatics in search.} The evaluation indicated a need to consider pragmatics to support a more natural conversational flow in search interfaces, a paradigm present in various natural language interfaces for data exploration~\cite{nl4dv,eviza}. Several (5 out of 12) participants typed a full query such as, ``\textit{which stocks went up in 2014?},'' followed by an underspecified query, ``\textit{what about 2015?}'' that would need to be interpreted in the context of the previous search history. P3 commented, ``\textit{I'm now used to just asking the next question assuming the system already knows what I mean. I expect the same [behavior] here too}.''

\pheading{Integrate trend search with visual analytical tools.} For supporting a more comprehensive data exploration, \toolname{} would need to be integrated into visual analysis tools that support a wider range of analytical inquiry. For example, three participants wanted to see categories of stock that did poorly or compare their relative performance with each other, as noted by P1: ``\textit{while I saw a bunch of stock tanking around the same time, I'd like to bucket them into a multi-line chart by tech vs. retail stock to get a better understanding of the trend patterns}.''

\pheading{Provide external knowledge that gives context around a certain trend pattern.} A deeper understanding of any trend necessitates an awareness of the contextual information surrounding it. While the search results in \toolname{} indicated a certain trend pattern, external knowledge can provide the \textit{why} and \textit{how} behind such a pattern, such as augmenting data with external information from knowledge graphs and web corpora~\cite{CashmanXDHLHGEC21}. 8 out of 12 participants expressed a need for including additional context beyond what is in the underlying data to enable them to make informed decisions and assess how the trends are influenced by external events. P4 stated, ``\textit{I can see what is the trend, but I'm curious to know more. Why did that stock suddenly go bust at that time? Feels like I want a Google button right next to that sharp drop}.''

\pheading{Support control over time granularity as well as fuzzy time concepts.} Finally, \toolname{} returns trend results by year, and participants expressed the need for more flexibility in exploring trends with varying granularity, from specific dates to weekly, monthly, and quarterly views, as well as fuzzy temporal descriptors, like ``stocks that remained stagnant for an extended period.'' P5 wanted to see the ``\textit{fluctuating trend quarter-by-quarter to check if there's anything seasonal going on there}.'' Computing trend patterns at different levels of time granularity could support more nuanced analyses and expose patterns such as seasonal variations or temporal outliers.

%% file: sections/6-future-work.tex
\section{Future Directions for Semantic Trends Dataset and its Applications}

In this work, we demonstrate the utility of a labeled dataset of semantic trends and their properties by implementing \toolname{} to help search and discover trends in line charts. However, we envision multiple directions for future work to utilize our dataset and contributions to go beyond the current capabilities of \toolname{}.

\subsection{Extending Semantic Trends Dataset}

One future direction for extending our dataset is to support users searching for more global descriptions of time series data behavior over a longer period of time (e.g., a user may want to search for when a certain stock was ``volatile'' versus ``consistent''). We could also leverage the fact that a conceptual duality exists between event sequences and global descriptors. For instance, one way of expressing that a stock is ``volatile'' is to describe the price as going up and down repeatedly, which could be represented by a sequence of trend events comprising ``up'' and ``down'' events.

LLMs, when trained on domain-specific corpora, can also help identify and label trends that are specifically meaningful to a particular domain. For instance, rather than identifying a trend as simply ``positive'' or ``negative,'' LLMs could be used to discern and label subtleties such as ``bullish,'' ``stagnant,'' and ``bearish'' for market trends, ``growth,'' ``plateau,'' and ``decline'' for financial data, or ``increasing,'' ``steady,'' and ``falling'' for weather information. LLMs also show promise in labeling trends at different temporal scales based on the data domain. We hence think it would be worthwhile to explore leveraging LLMs to label trends across different temporal scales -- for example, a short-term ``daily rally,'' a medium-term ``monthly correction,'' or a long-term ``annual growth trajectory.'' 

While our work introduces a search tool for trends, there are still opportunities with respect to data presentation (e.g., text interfaces, text summaries, accessibility tools, etc.).  LLMs could be employed to craft a narrative from a set of quantitative semantic observations such as those depicted in Figures \ref{fig:exp1_only_angle_labeling}, \ref{fig:exp2_only_angle_labeling}, and \ref{fig:exp3_shape_assignments}. 
For example, in Figure \ref{fig:gpt_narrative}, we asked GPT-4 to respect the specific trend labels but, if appropriate, to aggregate several events with labels like ``fluctuation.''  This ability to ``smooth out'' a set of events into a single engaging narrative could provide alternative ways to consume quantitative semantic observations. However, given the possibility of LLMs hallucinating in their responses, future work should also consider methods of checking for hallucinations in LLM-generated narratives and exposition text~\cite{zuccon:2023}.

\definecolor{gpt_narrative_figure_1}{RGB}{98,0,0}
\definecolor{gpt_narrative_figure_2}{RGB}{56,118,29}
\definecolor{gpt_narrative_figure_3}{RGB}{0,0,255}
\definecolor{gpt_narrative_figure_4}{RGB}{255,0,0}
\definecolor{gpt_narrative_figure_5}{RGB}{153,0,255}
\begin{figure}
    \centering
    \includegraphics[trim=0 100 0 0, clip, width=\columnwidth]{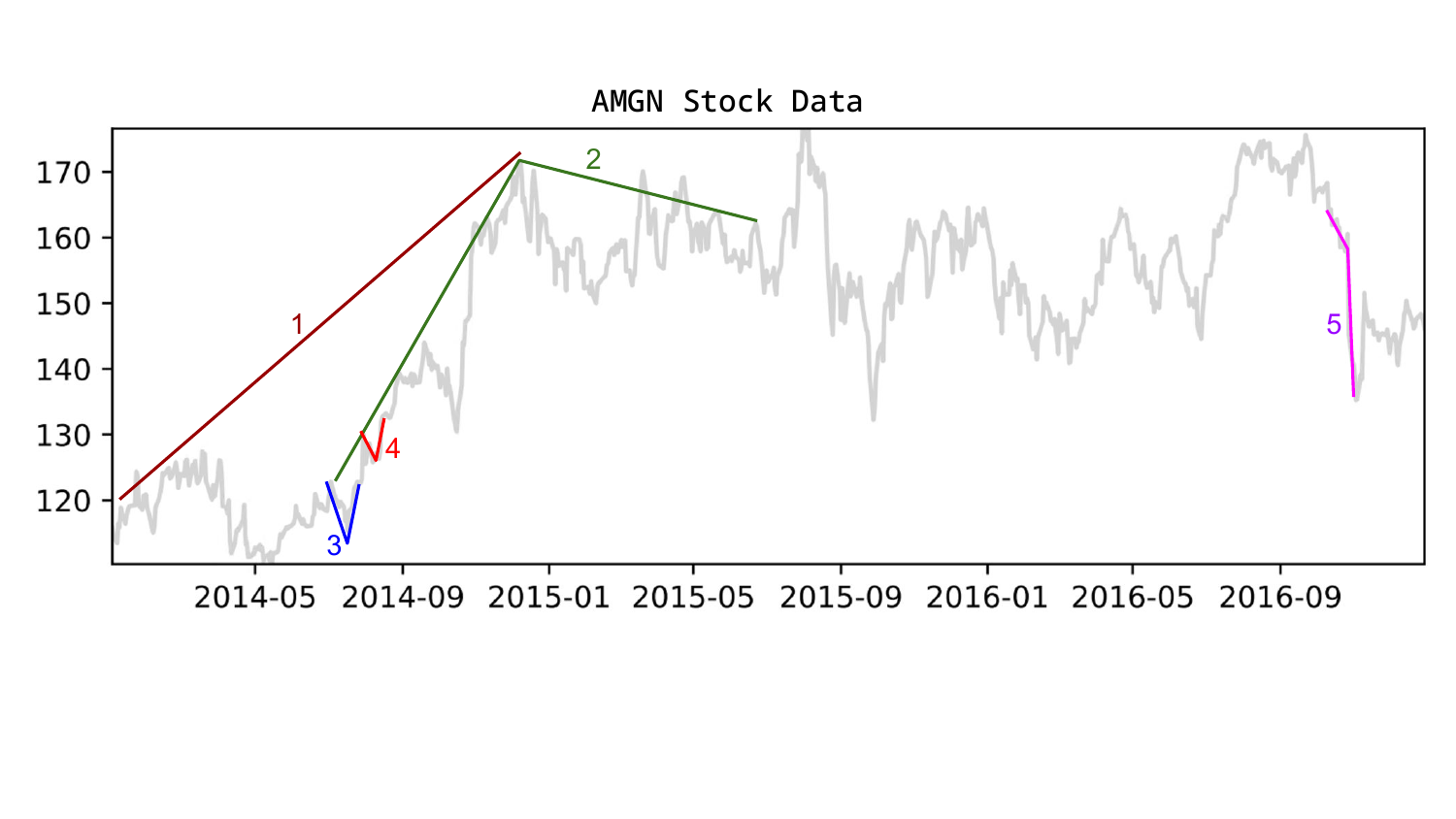}
    \caption{GPT-generated narrative based on quantitative semantic observations. Line color is for clarity only. (Numbers) refer to trend identifiers.
    \textit{``In the realm of biotechnology stocks, AMGN exhibited a distinct pattern of ups and downs. The stock started 2014 by \textcolor{gpt_narrative_figure_1}{\textup{(1)}} slowly expanding, growing by 32\% until December of that year. This upward movement evolved into a \textcolor{gpt_narrative_figure_2}{\textup{(2)}} hump shape that continued until July 2015, marking a 20\% increase from its starting point in mid-2014. The larger events were interspersed with smaller but still noteworthy fluctuations. For instance, the stock experienced a \textcolor{gpt_narrative_figure_3}{\textup{(3)}} rebound in July 2014, followed by a \textcolor{gpt_narrative_figure_4}{\textup{(4)}} lull that extended into September. Towards the end of 2016, the stock went through a \textcolor{gpt_narrative_figure_5}{\textup{(5)}} cliff-like decline of 17\%, a sharp contrast to its earlier growth. These smaller events added intricate layers to AMGN’s overall performance, rendering it a stock of compelling dynamics.''}}
    \Description[A GPT-generated narrative based on quantitative semantic observations.  Five slopes and shapes are shown and described by ChatGPT in the main figure text.]{Five slopes and shapes are overlaid on top of the stock data.  Slope 1 is a long upward slope from January 2014 to December 2014. Slope 2 is a large slightly rotated roof-style shape that rose from July 2014 until December 2014 at which point it went back down again until July 2015.  Shapes 3 and 4 are small V-style shapes in July and August of 2014.  Shape 5 is a steep downward slope in October of 2016.  The ChatGPT-generated narrative of these shapes is related to the main figure text.}
    \label{fig:gpt_narrative}
\end{figure}

\subsection{Future Extensions of Trend Search Tools}
Our work also points to potential new modalities and approaches for trend querying. For instance, the underlying labeled dataset and subsequent search techniques can be integrated with sketch-based input to help bridge the semantic gap between visual specification and trend semantics. While \toolname{} incorporates domain-specific labels for describing trends in data, the labeled dataset could be applied to provide analysis and data narratives bespoke to that domain. Search tools and visual analytics tools can also employ distinct lexicons to provide more targeted data exploration, insight, and narrative generation. While \toolname{} currently supports search queries that involve trend descriptors and modifiers such as ``fast falling'' and ``slow rising,'' the tool can be augmented with additional corpora and knowledge, including LLMs, to offer interpretations of trends that are contextually relevant, such as the query, ``\textit{Do stocks always fall after they bounce twice?}''.

%% file: sections/7-conclusion.tex
\section{Conclusion}


Search systems have begun to support basic analytical intents when displaying data and charts in response to users' natural language queries. However, user workflows often expect more specific tasks than can be robustly handled by search tools, such as identifying relevant trends in temporal data. In this paper, we present a dataset of trend descriptor labels and associated quantified semantics and then employ this dataset in a search tool called \toolname{}, which supports diverse trend search intents. The tool utilizes custom scoring and ranking logic to return relevant results based on users' natural language queries. A preliminary evaluation of \toolname{} demonstrates that the tool is intuitive for finding trends in data, and the underlying quantifiable semantic trend labels provide relevant search results for various nuances of trend descriptors in the input queries. We hope that our publicly available semantic trend label dataset can enable future research in developing intelligent search interfaces that can understand and leverage precise quantified semantics and support users' increasingly diverse visual data analysis intents.

%% file: main.bbl

\begin{thebibliography}{63}


\ifx \showCODEN    \undefined \def \showCODEN     #1{\unskip}     \fi
\ifx \showDOI      \undefined \def \showDOI       #1{#1}\fi
\ifx \showISBNx    \undefined \def \showISBNx     #1{\unskip}     \fi
\ifx \showISBNxiii \undefined \def \showISBNxiii  #1{\unskip}     \fi
\ifx \showISSN     \undefined \def \showISSN      #1{\unskip}     \fi
\ifx \showLCCN     \undefined \def \showLCCN      #1{\unskip}     \fi
\ifx \shownote     \undefined \def \shownote      #1{#1}          \fi
\ifx \showarticletitle \undefined \def \showarticletitle #1{#1}   \fi
\ifx \showURL      \undefined \def \showURL       {\relax}        \fi
\providecommand\bibfield[2]{#2}
\providecommand\bibinfo[2]{#2}
\providecommand\natexlab[1]{#1}
\providecommand\showeprint[2][]{arXiv:#2}

\bibitem[ela(2023)]%
        {elastic}
 \bibinfo{year}{2023}\natexlab{}.
\newblock \bibinfo{title}{{E}lasticsearch}.
\newblock \bibinfo{howpublished}{\url{https://www.elastic.co/elasticsearch/}}.
\newblock


\bibitem[fla(2023)]%
        {flask}
 \bibinfo{year}{2023}\natexlab{}.
\newblock \bibinfo{title}{{F}lask}.
\newblock \bibinfo{howpublished}{\url{https://flask.palletsprojects.com/en/3.0.x/}}.
\newblock


\bibitem[ibm(2023)]%
        {ibmwatson}
 \bibinfo{year}{2023}\natexlab{}.
\newblock \bibinfo{title}{{IBM} {W}atson {A}nalytics}.
\newblock \bibinfo{howpublished}{\url{http://www.ibm.com/analytics/watson-analytics}}.
\newblock


\bibitem[rea(2023)]%
        {react}
 \bibinfo{year}{2023}\natexlab{}.
\newblock \bibinfo{title}{{R}eact}.
\newblock \bibinfo{howpublished}{\url{https://react.dev/}}.
\newblock


\bibitem[sus(2023)]%
        {sus}
 \bibinfo{year}{2023}\natexlab{}.
\newblock \bibinfo{title}{System {U}sability {S}cale ({SUS})}.
\newblock \bibinfo{howpublished}{\url{https://www.usability.gov/how-to-and-tools/methods/system-usability-scale.html}}.
\newblock


\bibitem[tab(2023)]%
        {tableau}
 \bibinfo{year}{2023}\natexlab{}.
\newblock \bibinfo{title}{Tableau {S}oftware}.
\newblock \bibinfo{howpublished}{\url{https://www.tableau.com}}.
\newblock


\bibitem[tho(2023)]%
        {thoughtspot}
 \bibinfo{year}{2023}\natexlab{}.
\newblock \bibinfo{title}{{T}hought{S}pot}.
\newblock \bibinfo{howpublished}{\url{http://www.thoughtspot.com}}.
\newblock


\bibitem[Airio et~al\mbox{.}(2004)]%
        {ciri}
\bibfield{author}{\bibinfo{person}{Eija Airio}, \bibinfo{person}{Kalervo J{\"a}rvelin}, \bibinfo{person}{Pirkko Saatsi}, \bibinfo{person}{Jaana Kek{\"a}l{\"a}inen}, {and} \bibinfo{person}{Sari Suomela}.} \bibinfo{year}{2004}\natexlab{}.
\newblock \bibinfo{booktitle}{\emph{CIRI - An Ontology-based Query Interface for Text Retrieval} (\bibinfo{edition}{1} ed.)}.
\newblock Number~20 in \bibinfo{series}{Publications of the Finnish Artificial Intelligence Society}. \bibinfo{publisher}{Finnish Artificial Intelligence Society}, \bibinfo{pages}{73--82}.
\newblock
\showISBNx{951-96735-55}


\bibitem[Amar et~al\mbox{.}(2005)]%
        {amar2005tasks}
\bibfield{author}{\bibinfo{person}{Robert Amar}, \bibinfo{person}{James Eagan}, {and} \bibinfo{person}{John Stasko}.} \bibinfo{year}{2005}\natexlab{}.
\newblock \showarticletitle{Low-level components of analytic activity in information visualization}. In \bibinfo{booktitle}{\emph{IEEE Symposium on Information Visualization, 2005. INFOVIS 2005.}} IEEE, \bibinfo{pages}{111--117}.
\newblock
\urldef\tempurl%
\url{https://doi.org/10.1109/INFVIS.2005.1532136}
\showDOI{\tempurl}


\bibitem[Bromley and Setlur(2023)]%
        {bromley2023difference}
\bibfield{author}{\bibinfo{person}{Dennis Bromley} {and} \bibinfo{person}{Vidya Setlur}.} \bibinfo{year}{2023}\natexlab{}.
\newblock \showarticletitle{What Is the Difference Between a Mountain and a Molehill? Quantifying Semantic Labeling of Visual Features in Line Charts}.
\newblock \bibinfo{journal}{\emph{IEEE Transactions on Visualization and Computer Graphics}} (\bibinfo{year}{2023}).
\newblock


\bibitem[Buscaldi et~al\mbox{.}(2005)]%
        {Buscaldi2005AWQ}
\bibfield{author}{\bibinfo{person}{D. Buscaldi}, \bibinfo{person}{Paolo Rosso}, {and} \bibinfo{person}{Emilio~Sanchis Arnal}.} \bibinfo{year}{2005}\natexlab{}.
\newblock \showarticletitle{A WordNet-based Query Expansion Method for Geographical Information Retrieval}. In \bibinfo{booktitle}{\emph{Conference and Labs of the Evaluation Forum}}.
\newblock


\bibitem[Cashman et~al\mbox{.}(2021)]%
        {CashmanXDHLHGEC21}
\bibfield{author}{\bibinfo{person}{Dylan Cashman}, \bibinfo{person}{Shenyu Xu}, \bibinfo{person}{Subhajit Das}, \bibinfo{person}{Florian Heimerl}, \bibinfo{person}{Cong Liu}, \bibinfo{person}{Shah~Rukh Humayoun}, \bibinfo{person}{Michael Gleicher}, \bibinfo{person}{Alex Endert}, {and} \bibinfo{person}{Remco Chang}.} \bibinfo{year}{2021}\natexlab{}.
\newblock \showarticletitle{CAVA: A Visual Analytics System for Exploratory Columnar Data Augmentation Using Knowledge Graphs}.
\newblock \bibinfo{journal}{\emph{IEEE Transactions on Visualization and Computer Graphics}} \bibinfo{volume}{27}, \bibinfo{number}{2} (\bibinfo{year}{2021}), \bibinfo{pages}{1731--1741}.
\newblock
\urldef\tempurl%
\url{https://doi.org/10.1109/TVCG.2020.3030443}
\showDOI{\tempurl}


\bibitem[Chatfield(2004)]%
        {chatfield2004timeseries}
\bibfield{author}{\bibinfo{person}{Chris Chatfield}.} \bibinfo{year}{2004}\natexlab{}.
\newblock \bibinfo{booktitle}{\emph{The Analysis of Time Series: An Introduction} (\bibinfo{edition}{6th} ed.)}.
\newblock \bibinfo{publisher}{CRC Press}, \bibinfo{address}{Florida, US}.
\newblock


\bibitem[Cheng et~al\mbox{.}(2008)]%
        {falcons}
\bibfield{author}{\bibinfo{person}{Gong Cheng}, \bibinfo{person}{Weiyi Ge}, {and} \bibinfo{person}{Yuzhong Qu}.} \bibinfo{year}{2008}\natexlab{}.
\newblock \showarticletitle{Falcons: Searching and Browsing Entities on the Semantic Web}. In \bibinfo{booktitle}{\emph{Proceedings of the 17th International Conference on World Wide Web}} (Beijing, China) \emph{(\bibinfo{series}{WWW '08})}. \bibinfo{publisher}{Association for Computing Machinery}, \bibinfo{address}{New York, NY, USA}, \bibinfo{pages}{1101–1102}.
\newblock
\showISBNx{9781605580852}
\urldef\tempurl%
\url{https://doi.org/10.1145/1367497.1367676}
\showDOI{\tempurl}


\bibitem[Cimiano et~al\mbox{.}(2008)]%
        {cimiano:2008}
\bibfield{author}{\bibinfo{person}{Philipp Cimiano}, \bibinfo{person}{Peter Haase}, \bibinfo{person}{J\"{o}rg Heizmann}, \bibinfo{person}{Matthias Mantel}, {and} \bibinfo{person}{Rudi Studer}.} \bibinfo{year}{2008}\natexlab{}.
\newblock \showarticletitle{Towards Portable Natural Language Interfaces to Knowledge Bases - The Case of the ORAKEL System}.
\newblock \bibinfo{journal}{\emph{Data Knowl. Eng.}} \bibinfo{volume}{65}, \bibinfo{number}{2} (\bibinfo{date}{May} \bibinfo{year}{2008}), \bibinfo{pages}{325–354}.
\newblock
\showISSN{0169-023X}
\urldef\tempurl%
\url{https://doi.org/10.1016/j.datak.2007.10.007}
\showDOI{\tempurl}


\bibitem[Corby et~al\mbox{.}(2004)]%
        {corby:2004}
\bibfield{author}{\bibinfo{person}{Olivier Corby}, \bibinfo{person}{Rose Dieng-Kuntz}, {and} \bibinfo{person}{Catherine Faron-Zucker}.} \bibinfo{year}{2004}\natexlab{}.
\newblock \showarticletitle{Querying the Semantic Web with the CORESE search engine}.
\newblock \bibinfo{journal}{\emph{Proceedings of the 16th European Conference on Artificial Intelligence (ECAI 2004)}}, \bibinfo{pages}{705--709}.
\newblock


\bibitem[Damljanovic et~al\mbox{.}(2010)]%
        {Damljanovic2010NaturalLI}
\bibfield{author}{\bibinfo{person}{Danica Damljanovic}, \bibinfo{person}{Milan Agatonovic}, {and} \bibinfo{person}{Hamish Cunningham}.} \bibinfo{year}{2010}\natexlab{}.
\newblock \showarticletitle{Natural Language Interfaces to Ontologies: Combining Syntactic Analysis and Ontology-Based Lookup through the User Interaction}. In \bibinfo{booktitle}{\emph{Extended Semantic Web Conference}}.
\newblock


\bibitem[De~Oliveira and Levkowitz(2003)]%
        {de2003visual}
\bibfield{author}{\bibinfo{person}{MC~Ferreira De~Oliveira} {and} \bibinfo{person}{Haim Levkowitz}.} \bibinfo{year}{2003}\natexlab{}.
\newblock \showarticletitle{From Visual Data Exploration to Visual Data Mining: A Survey}.
\newblock \bibinfo{journal}{\emph{IEEE Transactions on Visualization and Computer Graphics}} \bibinfo{volume}{9}, \bibinfo{number}{3} (\bibinfo{year}{2003}), \bibinfo{pages}{378--394}.
\newblock


\bibitem[Dhamdhere et~al\mbox{.}(2017)]%
        {analyza}
\bibfield{author}{\bibinfo{person}{Kedar Dhamdhere}, \bibinfo{person}{Kevin~S. McCurley}, \bibinfo{person}{Ralfi Nahmias}, \bibinfo{person}{Mukund Sundararajan}, {and} \bibinfo{person}{Qiqi Yan}.} \bibinfo{year}{2017}\natexlab{}.
\newblock \showarticletitle{Analyza: Exploring Data with Conversation}. In \bibinfo{booktitle}{\emph{Proceedings of the 22nd International Conference on Intelligent User Interfaces}} \emph{(\bibinfo{series}{IUI 2017})}. \bibinfo{pages}{493--504}.
\newblock


\bibitem[Douglas and Peucker(1973)]%
        {douglas1973algorithms}
\bibfield{author}{\bibinfo{person}{David~H Douglas} {and} \bibinfo{person}{Thomas~K Peucker}.} \bibinfo{year}{1973}\natexlab{}.
\newblock \showarticletitle{Algorithms for the Reduction of the Number of Points Required to Represent a Digitized Line or its Caricature}.
\newblock \bibinfo{journal}{\emph{Cartographica: The International Journal for Geographic Information and Geovisualization}} \bibinfo{volume}{10}, \bibinfo{number}{2} (\bibinfo{year}{1973}), \bibinfo{pages}{112--122}.
\newblock


\bibitem[Fernandez et~al\mbox{.}(2008)]%
        {fernandez:2008}
\bibfield{author}{\bibinfo{person}{Miriam Fernandez}, \bibinfo{person}{Vanessa Lopez}, \bibinfo{person}{Marta Sabou}, \bibinfo{person}{Victoria Uren}, \bibinfo{person}{David Vallet}, \bibinfo{person}{Enrico Motta}, {and} \bibinfo{person}{Pablo Castells}.} \bibinfo{year}{2008}\natexlab{}.
\newblock \showarticletitle{Semantic Search Meets the Web}. In \bibinfo{booktitle}{\emph{2008 IEEE International Conference on Semantic Computing}}. \bibinfo{pages}{253--260}.
\newblock
\urldef\tempurl%
\url{https://doi.org/10.1109/ICSC.2008.52}
\showDOI{\tempurl}


\bibitem[Ferres et~al\mbox{.}(2006)]%
        {ferres2006igraph}
\bibfield{author}{\bibinfo{person}{Leo Ferres}, \bibinfo{person}{Avi Parush}, \bibinfo{person}{Zhihong Li}, \bibinfo{person}{Yandu Oppacher}, {and} \bibinfo{person}{Gitte Lindgaard}.} \bibinfo{year}{2006}\natexlab{}.
\newblock \showarticletitle{Representing and Querying Line Graphs in Natural Language: The \emph{iGraph} System}. In \bibinfo{booktitle}{\emph{Smart Graphics, 6th International Symposium, {SG} 2006, Vancouver, Canada, July 23-25, 2006, Proceedings}} \emph{(\bibinfo{series}{Lecture Notes in Computer Science}, Vol.~\bibinfo{volume}{4073})}, \bibfield{editor}{\bibinfo{person}{Andreas Butz}, \bibinfo{person}{Brian~D. Fisher}, \bibinfo{person}{Antonio Kr{\"{u}}ger}, {and} \bibinfo{person}{Patrick Olivier}} (Eds.). \bibinfo{publisher}{Springer}, \bibinfo{pages}{248--253}.
\newblock
\urldef\tempurl%
\url{https://doi.org/10.1007/11795018\_25}
\showDOI{\tempurl}


\bibitem[Finin et~al\mbox{.}(2005)]%
        {swoogle}
\bibfield{author}{\bibinfo{person}{Tim Finin}, \bibinfo{person}{Li Ding}, \bibinfo{person}{Rong Pan}, \bibinfo{person}{Anupam Joshi}, \bibinfo{person}{Pranam Kolari}, \bibinfo{person}{Akshay Java}, {and} \bibinfo{person}{Yun Peng}.} \bibinfo{year}{2005}\natexlab{}.
\newblock \showarticletitle{Swoogle: Searching for Knowledge on the Semantic Web}. In \bibinfo{booktitle}{\emph{Proceedings of the 20th National Conference on Artificial Intelligence - Volume 4}} (Pittsburgh, Pennsylvania) \emph{(\bibinfo{series}{AAAI'05})}. \bibinfo{publisher}{AAAI Press}, \bibinfo{pages}{1682–1683}.
\newblock
\showISBNx{157735236x}


\bibitem[Gao et~al\mbox{.}(2015)]%
        {datatone}
\bibfield{author}{\bibinfo{person}{Tong Gao}, \bibinfo{person}{Mira Dontcheva}, \bibinfo{person}{Eytan Adar}, \bibinfo{person}{Zhicheng Liu}, {and} \bibinfo{person}{Karrie~G. Karahalios}.} \bibinfo{year}{2015}\natexlab{}.
\newblock \showarticletitle{DataTone: Managing Ambiguity in Natural Language Interfaces for Data Visualization}. In \bibinfo{booktitle}{\emph{Proceedings of the 28th Annual ACM Symposium on User Interface Software Technology}} \emph{(\bibinfo{series}{UIST 2015})}. \bibinfo{publisher}{ACM}, \bibinfo{address}{New York, NY, USA}, \bibinfo{pages}{489--500}.
\newblock
\showISBNx{978-1-4503-3779-3}


\bibitem[Gruber(1993)]%
        {grubar:1993}
\bibfield{author}{\bibinfo{person}{Thomas~R. Gruber}.} \bibinfo{year}{1993}\natexlab{}.
\newblock \showarticletitle{A Translation Approach to Portable Ontology Specifications}.
\newblock \bibinfo{journal}{\emph{Knowledge Acquisition}} \bibinfo{volume}{5}, \bibinfo{number}{2} (\bibinfo{year}{1993}), \bibinfo{pages}{199--220}.
\newblock
\showISSN{1042-8143}
\urldef\tempurl%
\url{https://doi.org/10.1006/knac.1993.1008}
\showDOI{\tempurl}


\bibitem[Harth et~al\mbox{.}(2007)]%
        {Harth2007SWSEAB}
\bibfield{author}{\bibinfo{person}{A. Harth}, \bibinfo{person}{Aidan Hogan}, \bibinfo{person}{Renaud Delbru}, \bibinfo{person}{J{\"u}rgen Umbrich}, \bibinfo{person}{Se{\'a}n O'Riain}, {and} \bibinfo{person}{Stefan Decker}.} \bibinfo{year}{2007}\natexlab{}.
\newblock \showarticletitle{SWSE: Answers Before Links!}. In \bibinfo{booktitle}{\emph{Semantic Web Challenge}}.
\newblock


\bibitem[Heflin et~al\mbox{.}(2003)]%
        {Heflin2003SHOEAB}
\bibfield{author}{\bibinfo{person}{Jeff Heflin}, \bibinfo{person}{James~A. Hendler}, {and} \bibinfo{person}{Sean Luke}.} \bibinfo{year}{2003}\natexlab{}.
\newblock \showarticletitle{SHOE: A Blueprint for the Semantic Web}. In \bibinfo{booktitle}{\emph{Spinning the Semantic Web}}.
\newblock


\bibitem[Hochheiser and Shneiderman(2004)]%
        {Hochheiser2004DynamicQT}
\bibfield{author}{\bibinfo{person}{Harry Hochheiser} {and} \bibinfo{person}{Ben Shneiderman}.} \bibinfo{year}{2004}\natexlab{}.
\newblock \showarticletitle{Dynamic Query Tools for Time Series Data Sets: Timebox Widgets for Interactive Exploration}.
\newblock \bibinfo{journal}{\emph{Information Visualization}}  \bibinfo{volume}{3} (\bibinfo{year}{2004}), \bibinfo{pages}{1 -- 18}.
\newblock
\urldef\tempurl%
\url{https://api.semanticscholar.org/CorpusID:5628923}
\showURL{%
\tempurl}


\bibitem[Honnibal et~al\mbox{.}(2020)]%
        {spacy}
\bibfield{author}{\bibinfo{person}{Matthew Honnibal}, \bibinfo{person}{Ines Montani}, \bibinfo{person}{Sofie Van~Landeghem}, {and} \bibinfo{person}{Adriane Boyd}.} \bibinfo{year}{2020}\natexlab{}.
\newblock \showarticletitle{{spaCy: Industrial-strength Natural Language Processing in Python}}.
\newblock  (\bibinfo{year}{2020}).
\newblock
\urldef\tempurl%
\url{https://doi.org/10.5281/zenodo.1212303}
\showDOI{\tempurl}


\bibitem[Hoque et~al\mbox{.}(2022)]%
        {hoque2022cqa}
\bibfield{author}{\bibinfo{person}{E. Hoque}, \bibinfo{person}{P. Kavehzadeh}, {and} \bibinfo{person}{A. Masry}.} \bibinfo{year}{2022}\natexlab{}.
\newblock \showarticletitle{Chart Question Answering: State of the Art and Future Directions}.
\newblock \bibinfo{journal}{\emph{Computer Graphics Forum}} \bibinfo{volume}{41}, \bibinfo{number}{3} (\bibinfo{year}{2022}), \bibinfo{pages}{555--572}.
\newblock
\urldef\tempurl%
\url{https://doi.org/10.1111/cgf.14573}
\showDOI{\tempurl}


\bibitem[Hoque et~al\mbox{.}(2017)]%
        {hoque2017applying}
\bibfield{author}{\bibinfo{person}{Enamul Hoque}, \bibinfo{person}{Vidya Setlur}, \bibinfo{person}{Melanie Tory}, {and} \bibinfo{person}{Isaac Dykeman}.} \bibinfo{year}{2017}\natexlab{}.
\newblock \showarticletitle{Applying Pragmatics Principles for Interaction with Visual Analytics}.
\newblock \bibinfo{journal}{\emph{IEEE Transactions on Visualization and Computer Graphics}} \bibinfo{volume}{24}, \bibinfo{number}{1} (\bibinfo{year}{2017}), \bibinfo{pages}{309--318}.
\newblock


\bibitem[Kaufmann et~al\mbox{.}(2006)]%
        {kaufmann:2006}
\bibfield{author}{\bibinfo{person}{Esther Kaufmann}, \bibinfo{person}{Abraham Bernstein}, {and} \bibinfo{person}{Renato Zumstein}.} \bibinfo{year}{2006}\natexlab{}.
\newblock \showarticletitle{Querix: A Natural Language Interface to Query Ontologies Based on Clarification Dialogs}.
\newblock  (\bibinfo{date}{01} \bibinfo{year}{2006}).
\newblock


\bibitem[Kim et~al\mbox{.}(2021)]%
        {kim2021captions}
\bibfield{author}{\bibinfo{person}{Dae~Hyun Kim}, \bibinfo{person}{Vidya Setlur}, {and} \bibinfo{person}{Maneesh Agrawala}.} \bibinfo{year}{2021}\natexlab{}.
\newblock \showarticletitle{Towards Understanding How Readers Integrate Charts and Captions: A Case Study with Line Charts}. In \bibinfo{booktitle}{\emph{Proceedings of the 2021 CHI Conference on Human Factors in Computing Systems}} (Yokohama, Japan) \emph{(\bibinfo{series}{CHI '21})}. \bibinfo{publisher}{Association for Computing Machinery}, \bibinfo{address}{New York, NY, USA}, Article \bibinfo{articleno}{610}, \bibinfo{numpages}{11}~pages.
\newblock
\showISBNx{9781450380966}
\urldef\tempurl%
\url{https://doi.org/10.1145/3411764.3445443}
\showDOI{\tempurl}


\bibitem[Kincaid and Lam(2006)]%
        {kincaid2006line}
\bibfield{author}{\bibinfo{person}{Robert Kincaid} {and} \bibinfo{person}{Heidi Lam}.} \bibinfo{year}{2006}\natexlab{}.
\newblock \showarticletitle{Line Graph Explorer: Scalable Display of Line Graphs Using Focus+Context}. In \bibinfo{booktitle}{\emph{Proceedings of the Working Conference on Advanced Visual Interfaces}}. \bibinfo{pages}{404--411}.
\newblock


\bibitem[Klein and Manning(2003)]%
        {klein-manning-2003-accurate}
\bibfield{author}{\bibinfo{person}{Dan Klein} {and} \bibinfo{person}{Christopher~D. Manning}.} \bibinfo{year}{2003}\natexlab{}.
\newblock \showarticletitle{Accurate Unlexicalized Parsing}. In \bibinfo{booktitle}{\emph{Proceedings of the 41st Annual Meeting of the Association for Computational Linguistics}}. \bibinfo{publisher}{Association for Computational Linguistics}, \bibinfo{address}{Sapporo, Japan}, \bibinfo{pages}{423--430}.
\newblock
\urldef\tempurl%
\url{https://doi.org/10.3115/1075096.1075150}
\showDOI{\tempurl}


\bibitem[Lee et~al\mbox{.}(2020)]%
        {lee2019sensemaking}
\bibfield{author}{\bibinfo{person}{Doris Jung-Lin Lee}, \bibinfo{person}{John Lee}, \bibinfo{person}{Tarique Siddiqui}, \bibinfo{person}{Jaewoo Kim}, \bibinfo{person}{Karrie Karahalios}, {and} \bibinfo{person}{Aditya Parameswaran}.} \bibinfo{year}{2020}\natexlab{}.
\newblock \showarticletitle{You Can't Always Sketch What You Want: Understanding Sensemaking in Visual Query Systems}.
\newblock \bibinfo{journal}{\emph{IEEE Transactions on Visualization and Computer Graphics}} \bibinfo{volume}{26}, \bibinfo{number}{1} (\bibinfo{year}{2020}), \bibinfo{pages}{1267--1277}.
\newblock
\urldef\tempurl%
\url{https://doi.org/10.1109/TVCG.2019.2934666}
\showDOI{\tempurl}


\bibitem[Lei et~al\mbox{.}(2006)]%
        {Lei2006SemSearchAS}
\bibfield{author}{\bibinfo{person}{Yuangui Lei}, \bibinfo{person}{Victoria~S. Uren}, {and} \bibinfo{person}{Enrico Motta}.} \bibinfo{year}{2006}\natexlab{}.
\newblock \showarticletitle{SemSearch: A Search Engine for the Semantic Web}. In \bibinfo{booktitle}{\emph{International Conference Knowledge Engineering and Knowledge Management}}.
\newblock


\bibitem[L{\'o}pez et~al\mbox{.}(2005)]%
        {Lpez2005AquaLogAO}
\bibfield{author}{\bibinfo{person}{V. L{\'o}pez}, \bibinfo{person}{Michele Pasin}, {and} \bibinfo{person}{Enrico Motta}.} \bibinfo{year}{2005}\natexlab{}.
\newblock \showarticletitle{AquaLog: An Ontology-Portable Question Answering System for the Semantic Web}. In \bibinfo{booktitle}{\emph{Extended Semantic Web Conference}}.
\newblock


\bibitem[Lopez et~al\mbox{.}(2006)]%
        {lopez:2006}
\bibfield{author}{\bibinfo{person}{Vanessa Lopez}, \bibinfo{person}{Marta Sabou}, {and} \bibinfo{person}{Enrico Motta}.} \bibinfo{year}{2006}\natexlab{}.
\newblock \showarticletitle{PowerMap: Mapping the Real Semantic Web on the Fly}. In \bibinfo{booktitle}{\emph{Proceedings of the 5th International Conference on The Semantic Web}} (Athens, GA) \emph{(\bibinfo{series}{ISWC'06})}. \bibinfo{publisher}{Springer-Verlag}, \bibinfo{address}{Berlin, Heidelberg}, \bibinfo{pages}{414–427}.
\newblock
\showISBNx{3540490299}
\urldef\tempurl%
\url{https://doi.org/10.1007/11926078_30}
\showDOI{\tempurl}


\bibitem[Microsoft(2023)]%
        {powerbi}
\bibfield{author}{\bibinfo{person}{Microsoft}.} \bibinfo{year}{2023}\natexlab{}.
\newblock \bibinfo{title}{Microsoft PowerBI}.
\newblock \bibinfo{howpublished}{\url{https://powerbi.microsoft.com/}}.
\newblock


\bibitem[Miller(1995)]%
        {miller1995wordnet}
\bibfield{author}{\bibinfo{person}{George~A Miller}.} \bibinfo{year}{1995}\natexlab{}.
\newblock \showarticletitle{WordNet: A Lexical Database for English}.
\newblock \bibinfo{journal}{\emph{Commun. ACM}} \bibinfo{volume}{38}, \bibinfo{number}{11} (\bibinfo{year}{1995}), \bibinfo{pages}{39--41}.
\newblock


\bibitem[Miranda et~al\mbox{.}(2018)]%
        {Miranda2018TimeLA}
\bibfield{author}{\bibinfo{person}{F{\'a}bio Miranda}, \bibinfo{person}{Marcos Lage}, \bibinfo{person}{Harish Doraiswamy}, \bibinfo{person}{Charlie Mydlarz}, \bibinfo{person}{Justin Salamon}, \bibinfo{person}{Yitzchak~David Lockerman}, \bibinfo{person}{Juliana Freire}, {and} \bibinfo{person}{Cl{\'a}udio~T. Silva}.} \bibinfo{year}{2018}\natexlab{}.
\newblock \showarticletitle{Time Lattice: A Data Structure for the Interactive Visual Analysis of Large Time Series}.
\newblock \bibinfo{journal}{\emph{Computer Graphics Forum}}  \bibinfo{volume}{37} (\bibinfo{year}{2018}).
\newblock
\urldef\tempurl%
\url{https://api.semanticscholar.org/CorpusID:51873868}
\showURL{%
\tempurl}


\bibitem[Moldovan and Mihalcea(2000)]%
        {moldovan:2000}
\bibfield{author}{\bibinfo{person}{Dan~I. Moldovan} {and} \bibinfo{person}{Rada Mihalcea}.} \bibinfo{year}{2000}\natexlab{}.
\newblock \showarticletitle{Using WordNet and Lexical Operators to Improve Internet Searches}.
\newblock \bibinfo{journal}{\emph{IEEE Internet Computing}} \bibinfo{volume}{4}, \bibinfo{number}{1} (\bibinfo{date}{jan} \bibinfo{year}{2000}), \bibinfo{pages}{34–43}.
\newblock
\showISSN{1089-7801}
\urldef\tempurl%
\url{https://doi.org/10.1109/4236.815847}
\showDOI{\tempurl}


\bibitem[Narechania et~al\mbox{.}(2021)]%
        {nl4dv}
\bibfield{author}{\bibinfo{person}{Arpit Narechania}, \bibinfo{person}{Arjun Srinivasan}, {and} \bibinfo{person}{John Stasko}.} \bibinfo{year}{2021}\natexlab{}.
\newblock \showarticletitle{NL4DV: A Toolkit for Generating Analytic Specifications for Data Visualization from Natural Language Queries}.
\newblock \bibinfo{journal}{\emph{IEEE Transactions on Visualization and Computer Graphics}} \bibinfo{volume}{27}, \bibinfo{number}{2} (\bibinfo{year}{2021}), \bibinfo{pages}{369--379}.
\newblock
\urldef\tempurl%
\url{https://doi.org/10.1109/TVCG.2020.3030378}
\showDOI{\tempurl}


\bibitem[OpenAI(2023)]%
        {openai2023gpt}
\bibfield{author}{\bibinfo{person}{OpenAI}.} \bibinfo{year}{2023}\natexlab{}.
\newblock \showarticletitle{GPT-4 technical report}.
\newblock \bibinfo{journal}{\emph{arXiv}} (\bibinfo{year}{2023}), \bibinfo{pages}{2303--08774}.
\newblock


\bibitem[Oren et~al\mbox{.}(2008)]%
        {oren:2008}
\bibfield{author}{\bibinfo{person}{Eyal Oren}, \bibinfo{person}{Christophe Gu\'{e}ret}, {and} \bibinfo{person}{Stefan Schlobach}.} \bibinfo{year}{2008}\natexlab{}.
\newblock \showarticletitle{Anytime Query Answering in RDF through Evolutionary Algorithms}. In \bibinfo{booktitle}{\emph{Proceedings of the 7th International Conference on The Semantic Web}} (Karlsruhe, Germany) \emph{(\bibinfo{series}{ISWC '08})}. \bibinfo{publisher}{Springer-Verlag}, \bibinfo{address}{Berlin, Heidelberg}, \bibinfo{pages}{98–113}.
\newblock
\showISBNx{9783540885634}
\urldef\tempurl%
\url{https://doi.org/10.1007/978-3-540-88564-1_7}
\showDOI{\tempurl}


\bibitem[Pedregosa et~al\mbox{.}(2011)]%
        {scikit-learn}
\bibfield{author}{\bibinfo{person}{F. Pedregosa}, \bibinfo{person}{G. Varoquaux}, \bibinfo{person}{A. Gramfort}, \bibinfo{person}{V. Michel}, \bibinfo{person}{B. Thirion}, \bibinfo{person}{O. Grisel}, \bibinfo{person}{M. Blondel}, \bibinfo{person}{P. Prettenhofer}, \bibinfo{person}{R. Weiss}, \bibinfo{person}{V. Dubourg}, \bibinfo{person}{J. Vanderplas}, \bibinfo{person}{A. Passos}, \bibinfo{person}{D. Cournapeau}, \bibinfo{person}{M. Brucher}, \bibinfo{person}{M. Perrot}, {and} \bibinfo{person}{E. Duchesnay}.} \bibinfo{year}{2011}\natexlab{}.
\newblock \showarticletitle{Scikit-learn: Machine Learning in {P}ython}.
\newblock \bibinfo{journal}{\emph{Journal of Machine Learning Research}}  \bibinfo{volume}{12} (\bibinfo{year}{2011}), \bibinfo{pages}{2825--2830}.
\newblock


\bibitem[{PostgreSQL Global Development Group}(2023)]%
        {postgres}
\bibfield{author}{\bibinfo{person}{{PostgreSQL Global Development Group}}.} \bibinfo{year}{2023}\natexlab{}.
\newblock \bibinfo{title}{PostgreSQL}.
\newblock
\newblock
\urldef\tempurl%
\url{https://www.postgresql.org/}
\showURL{%
\tempurl}


\bibitem[Setlur et~al\mbox{.}(2016)]%
        {eviza}
\bibfield{author}{\bibinfo{person}{Vidya Setlur}, \bibinfo{person}{Sarah~E. Battersby}, \bibinfo{person}{Melanie Tory}, \bibinfo{person}{Rich Gossweiler}, {and} \bibinfo{person}{Angel~X. Chang}.} \bibinfo{year}{2016}\natexlab{}.
\newblock \showarticletitle{Eviza: A Natural Language Interface for Visual Analysis}. In \bibinfo{booktitle}{\emph{Proceedings of the 29th Annual Symposium on User Interface Software and Technology}} (Tokyo, Japan) \emph{(\bibinfo{series}{UIST 2016})}. \bibinfo{publisher}{ACM}, \bibinfo{address}{New York, NY, USA}, \bibinfo{pages}{365--377}.
\newblock
\showISBNx{978-1-4503-4189-9}


\bibitem[Setlur et~al\mbox{.}(2023)]%
        {olio}
\bibfield{author}{\bibinfo{person}{Vidya Setlur}, \bibinfo{person}{Andriy Kanyuka}, {and} \bibinfo{person}{Arjun Srinivasan}.} \bibinfo{year}{2023}\natexlab{}.
\newblock \showarticletitle{Olio: A Semantic Search Interface for Data Repositories}. In \bibinfo{booktitle}{\emph{Proceedings of the 36th Annual ACM Symposium on User Interface Software Technology}} (San Francisco, California) \emph{(\bibinfo{series}{UIST 2023})}. \bibinfo{publisher}{ACM}, \bibinfo{address}{New York, NY, USA}.
\newblock
\showISBNx{979-8-4007-0132-0}
\urldef\tempurl%
\url{https://doi.org/10.1145/3586183.3606806}
\showDOI{\tempurl}


\bibitem[Setlur et~al\mbox{.}(2019)]%
        {setlur2019inferencing}
\bibfield{author}{\bibinfo{person}{Vidya Setlur}, \bibinfo{person}{Melanie Tory}, {and} \bibinfo{person}{Alex Djalali}.} \bibinfo{year}{2019}\natexlab{}.
\newblock \showarticletitle{Inferencing Underspecified Natural Language Utterances in Visual Analysis}. In \bibinfo{booktitle}{\emph{Proceedings of the 24th International Conference on Intelligent User Interfaces}} (Marina del Rey, California) \emph{(\bibinfo{series}{IUI '19})}. \bibinfo{publisher}{Association for Computing Machinery}, \bibinfo{address}{New York, NY, USA}, \bibinfo{pages}{40--51}.
\newblock
\showISBNx{9781450362726}
\urldef\tempurl%
\url{https://doi.org/10.1145/3301275.3302270}
\showDOI{\tempurl}


\bibitem[Siddiqui et~al\mbox{.}(2016)]%
        {siddiqui2016zenvisage}
\bibfield{author}{\bibinfo{person}{Tarique Siddiqui}, \bibinfo{person}{Albert Kim}, \bibinfo{person}{John Lee}, \bibinfo{person}{Karrie Karahalios}, {and} \bibinfo{person}{Aditya Parameswaran}.} \bibinfo{year}{2016}\natexlab{}.
\newblock \showarticletitle{Effortless Data Exploration with Zenvisage: An Expressive and Interactive Visual Analytics System}.
\newblock \bibinfo{journal}{\emph{Proc. VLDB Endow.}} \bibinfo{volume}{10}, \bibinfo{number}{4} (\bibinfo{date}{nov} \bibinfo{year}{2016}), \bibinfo{pages}{457–468}.
\newblock
\showISSN{2150-8097}
\urldef\tempurl%
\url{https://doi.org/10.14778/3025111.3025126}
\showDOI{\tempurl}


\bibitem[Siddiqui et~al\mbox{.}(2021)]%
        {siddiqui2021shapesearch}
\bibfield{author}{\bibinfo{person}{Tarique Siddiqui}, \bibinfo{person}{Paul Luh}, \bibinfo{person}{Zesheng Wang}, \bibinfo{person}{Karrie Karahalios}, {and} \bibinfo{person}{Aditya~G. Parameswaran}.} \bibinfo{year}{2021}\natexlab{}.
\newblock \showarticletitle{From Sketching to Natural Language: Expressive Visual Querying for Accelerating Insight}.
\newblock \bibinfo{journal}{\emph{SIGMOD Rec.}} \bibinfo{volume}{50}, \bibinfo{number}{1} (\bibinfo{date}{jun} \bibinfo{year}{2021}), \bibinfo{pages}{51–58}.
\newblock
\showISSN{0163-5808}
\urldef\tempurl%
\url{https://doi.org/10.1145/3471485.3471498}
\showDOI{\tempurl}


\bibitem[Soukup and Davidson(2002)]%
        {soukup2002visual}
\bibfield{author}{\bibinfo{person}{Tom Soukup} {and} \bibinfo{person}{Ian Davidson}.} \bibinfo{year}{2002}\natexlab{}.
\newblock \bibinfo{booktitle}{\emph{Visual Data Mining: Techniques and Tools for Data Visualization and Mining}}.
\newblock \bibinfo{publisher}{John Wiley \& Sons}.
\newblock


\bibitem[Srihari and Li(1999)]%
        {Srihari1999InformationES}
\bibfield{author}{\bibinfo{person}{Rohini~K. Srihari} {and} \bibinfo{person}{W. Li}.} \bibinfo{year}{1999}\natexlab{}.
\newblock \showarticletitle{Information Extraction Supported Question Answering}. In \bibinfo{booktitle}{\emph{Text Retrieval Conference}}.
\newblock


\bibitem[Srinivasan and Stasko(2018)]%
        {orko}
\bibfield{author}{\bibinfo{person}{Arjun Srinivasan} {and} \bibinfo{person}{John Stasko}.} \bibinfo{year}{2018}\natexlab{}.
\newblock \showarticletitle{Orko: Facilitating Multimodal Interaction for Visual Exploration and Analysis of Networks}.
\newblock \bibinfo{journal}{\emph{IEEE Transactions on Visualization and Computer Graphics}} \bibinfo{volume}{24}, \bibinfo{number}{1} (\bibinfo{year}{2018}), \bibinfo{pages}{511--521}.
\newblock


\bibitem[Tunnicliffe~Wilson(2016)]%
        {tunnicliffe:2016}
\bibfield{author}{\bibinfo{person}{Granville Tunnicliffe~Wilson}.} \bibinfo{year}{2016}\natexlab{}.
\newblock \showarticletitle{Time Series Analysis: Forecasting and Control,5th Edition, by George E. P. Box, Gwilym M. Jenkins, Gregory C. Reinsel and Greta M. Ljung, 2015. Published by John Wiley and Sons Inc., Hoboken, New Jersey, pp. 712. ISBN: 978-1-118-67502-1}.
\newblock \bibinfo{journal}{\emph{Journal of Time Series Analysis}}  \bibinfo{volume}{37} (\bibinfo{date}{03} \bibinfo{year}{2016}).
\newblock
\urldef\tempurl%
\url{https://doi.org/10.1111/jtsa.12194}
\showDOI{\tempurl}


\bibitem[Wang et~al\mbox{.}(2007)]%
        {panto}
\bibfield{author}{\bibinfo{person}{Chong Wang}, \bibinfo{person}{Miao Xiong}, \bibinfo{person}{Qi Zhou}, {and} \bibinfo{person}{Yong Yu}.} \bibinfo{year}{2007}\natexlab{}.
\newblock \showarticletitle{PANTO: A Portable Natural Language Interface to Ontologies}. In \bibinfo{booktitle}{\emph{Proceedings of the 4th European Conference on The Semantic Web: Research and Applications}} (Innsbruck, Austria) \emph{(\bibinfo{series}{ESWC '07})}. \bibinfo{publisher}{Springer-Verlag}, \bibinfo{address}{Berlin, Heidelberg}, \bibinfo{pages}{473–487}.
\newblock
\showISBNx{9783540726661}
\urldef\tempurl%
\url{https://doi.org/10.1007/978-3-540-72667-8_34}
\showDOI{\tempurl}


\bibitem[Wong and Mooney(2006)]%
        {wong-mooney-2006-learning}
\bibfield{author}{\bibinfo{person}{Yuk~Wah Wong} {and} \bibinfo{person}{Raymond Mooney}.} \bibinfo{year}{2006}\natexlab{}.
\newblock \showarticletitle{Learning for Semantic Parsing with Statistical Machine Translation}. In \bibinfo{booktitle}{\emph{Proceedings of the Human Language Technology Conference of the {NAACL}, Main Conference}}. \bibinfo{publisher}{Association for Computational Linguistics}, \bibinfo{address}{New York City, USA}, \bibinfo{pages}{439--446}.
\newblock
\urldef\tempurl%
\url{https://aclanthology.org/N06-1056}
\showURL{%
\tempurl}


\bibitem[Yu and Silva(2020)]%
        {flowsense}
\bibfield{author}{\bibinfo{person}{B. Yu} {and} \bibinfo{person}{C.~T. Silva}.} \bibinfo{year}{2020}\natexlab{}.
\newblock \showarticletitle{FlowSense: A Natural Language Interface for Visual Data Exploration within a Dataflow System}.
\newblock \bibinfo{journal}{\emph{IEEE Transactions on Visualization and Computer Graphics}} \bibinfo{volume}{26}, \bibinfo{number}{01} (\bibinfo{date}{jan} \bibinfo{year}{2020}), \bibinfo{pages}{1--11}.
\newblock
\showISSN{1941-0506}
\urldef\tempurl%
\url{https://doi.org/10.1109/TVCG.2019.2934668}
\showDOI{\tempurl}


\bibitem[Zenz et~al\mbox{.}(2009)]%
        {zenz:2009}
\bibfield{author}{\bibinfo{person}{Gideon Zenz}, \bibinfo{person}{Xuan Zhou}, \bibinfo{person}{Enrico Minack}, \bibinfo{person}{Wolf Siberski}, {and} \bibinfo{person}{Wolfgang Nejdl}.} \bibinfo{year}{2009}\natexlab{}.
\newblock \showarticletitle{From Keywords to Semantic Queries-Incremental Query Construction on the Semantic Web}.
\newblock \bibinfo{journal}{\emph{Web Semant.}} \bibinfo{volume}{7}, \bibinfo{number}{3} (\bibinfo{date}{Sep} \bibinfo{year}{2009}), \bibinfo{pages}{166–176}.
\newblock
\showISSN{1570-8268}
\urldef\tempurl%
\url{https://doi.org/10.1016/j.websem.2009.07.005}
\showDOI{\tempurl}


\bibitem[Zhao et~al\mbox{.}(2022)]%
        {kd-box}
\bibfield{author}{\bibinfo{person}{Yue Zhao}, \bibinfo{person}{Yunhai Wang}, \bibinfo{person}{Jian Zhang}, \bibinfo{person}{Chi-Wing Fu}, \bibinfo{person}{Mingliang Xu}, {and} \bibinfo{person}{Dominik Moritz}.} \bibinfo{year}{2022}\natexlab{}.
\newblock \showarticletitle{KD-Box: Line-segment-based KD-tree for Interactive Exploration of Large-scale Time-Series Data}.
\newblock \bibinfo{journal}{\emph{IEEE Transactions on Visualization and Computer Graphics}} \bibinfo{volume}{28}, \bibinfo{number}{1} (\bibinfo{year}{2022}), \bibinfo{pages}{890--900}.
\newblock
\urldef\tempurl%
\url{https://doi.org/10.1109/TVCG.2021.3114865}
\showDOI{\tempurl}


\bibitem[Zuccon et~al\mbox{.}(2023)]%
        {zuccon:2023}
\bibfield{author}{\bibinfo{person}{Guido Zuccon}, \bibinfo{person}{Bevan Koopman}, {and} \bibinfo{person}{Razia Shaik}.} \bibinfo{year}{2023}\natexlab{}.
\newblock \showarticletitle{ChatGPT Hallucinates when Attributing Answers}. In \bibinfo{booktitle}{\emph{Proceedings of the Annual International ACM SIGIR Conference on Research and Development in Information Retrieval in the Asia Pacific Region}} (Beijing, China) \emph{(\bibinfo{series}{SIGIR-AP '23})}. \bibinfo{publisher}{Association for Computing Machinery}, \bibinfo{address}{New York, NY, USA}, \bibinfo{pages}{46–51}.
\newblock
\showISBNx{9798400704086}
\urldef\tempurl%
\url{https://doi.org/10.1145/3624918.3625329}
\showDOI{\tempurl}


\end{thebibliography}
